\documentclass[preview,authoryear,3p,onecolumn]{elsarticle}
\usepackage{float,graphicx}
\usepackage{amssymb,amsmath,amsfonts,amstext,bm}
\usepackage{ifpdf,lineno,natbib,url}
\usepackage[colorlinks=true,breaklinks=true,linkcolor={blue}]{hyperref}
\usepackage{epstopdf}
\usepackage{graphics}
\usepackage{color}
\usepackage{psfrag}
\usepackage{calc}
\usepackage[english]{algorithm2e}
\usepackage{textcomp}
\usepackage{epstopdf}
\usepackage{accents}
\graphicspath{{./figures/}}
\numberwithin{equation}{section}
\usepackage{lineno}

\usepackage{empheq} %for the braces around equations
\usepackage{mathtools} %for the dcase : the brace with different options in display mode, otherwise use cases of amsmath but it is not displaystyle. Also for the broken denominator of the long expression of recombination-generation
\usepackage{color,soul} 
\usepackage{bibentry} % Temporary for not printing the bilbiography at the end with the command nobibliography
\usepackage{enumitem}
\usepackage{stmaryrd} %jump notation
\usepackage{multirow} %multirow table
\usepackage[font=normalsize]{subfig}

\usepackage{color}
\makeatletter
\def\mathcolor#1#{\@mathcolor{#1}}
\def\@mathcolor#1#2#3{%
  \protect\leavevmode
  \begingroup
    \color#1{#2}#3%
  \endgroup
}
\makeatother
\newcommand{\cb}[1]{\mathcolor{red}{#1}}
\usepackage{etoolbox}
\makeatletter
\def\@mkboth#1#2{}
\newlength\appendixwidth
\preto\appendix{\addtocontents{toc}{\protect\patchl@section}}
\newcommand{\patchl@section}{%
    \settowidth{\appendixwidth}{\textbf{Appendix }}%
    \addtolength{\appendixwidth}{1.5em}%
    \patchcmd{\l@section}{1.5em}{\appendixwidth}{}{\ddt}%
}
\makeatother

\newcommand*\oF{\overline{F}}
\newcommand*\onu{\overline{\nu}}
\newcommand*\ok{\overline{\kappa}}
\newcommand*\okp{\overline{\kappa}_p}
\newcommand*\oa{\overline{\alpha}}
\newcommand*\ob{\overline{\beta}}
\newcommand\rz{\overset{0}{\rho}\vphantom{\rho}}

\newcommand\rhz{\overset{o}{\hat {\rho}}\vphantom{\rho}}
\newcommand\vz{\overset{0}{V}\vphantom{V}}

\newcommand\xz{\overset{0}{x}\vphantom{x}}

\newcommand{\me}{\mathrm{e}}
\newcommand\dc{^\circ \mathrm{C}}
\newcommand\rt{\tilde{\rho}}

\newcommand\rtz{\overset{0}{\tilde{\rho}}}
\newcommand\rtzp{\overset{0}{\tilde{\rho}'}}

\newcommand{\drt}{\delta \rt}
\newcommand{\dx}{\delta x}
\newcommand{\ddx}{\delta \dot{x}}
\newcommand{\drh}{\delta \hat{\rho}}

\newcommand{\dxh}{\delta \hat{x}}
\newcommand{\ha}{\hat{A}}
\newcommand{\hb}{\hat{B}}
\newcommand{\re}{\mathrm{Re}}

\newcommand{\drb}{\delta \check{\bm{\rho}}}

\newcommand{\ie}{i.e.}

\newcommand{\calE}{\mathcal{E}}

\usepackage[normalem]{ulem}

\begin{document}
\bibliographystyle{elsarticle-harv}

\begin{frontmatter}

%===================================================================
% Main document
%===================================================================

\title{Revisiting step instabilities on crystal surfaces. Part II: General theory.} %

\author[lms,lpicm,eth]{L.~Guin\corref{ca}}
\ead{laguin@ethz.ch}
\cortext[ca]{Corresponding author}
\author[lms,meca]{M.~E.~Jabbour}
\author[ladhyx,onera]{L. Shaabani-Ardali}
\author[lms,meca,um]{N.~Triantafyllidis}
\bigskip\bigskip

\address[lms]{LMS, \'{E}cole polytechnique, CNRS, Institut Polytechnique de Paris, 91128 Palaiseau, France}
\address[lpicm]{LPICM, \'{E}cole polytechnique, CNRS, Institut Polytechnique de Paris, 91128 Palaiseau, France}
\address[eth]{Mechanics \& Materials Lab, Department of Mechanical and Process Engineering, ETH Zürich, 8092 Zürich, Switzerland}
\address[meca]{D\'{e}partement de M\'{e}canique, \'{E}cole polytechnique, 91128 Palaiseau, France} 
\address[um]{Departments of Aerospace Engineering \& Mechanical Engineering (Emeritus)\\
The University of Michigan, Ann Arbor, MI 48109-2140, USA}
\address[ladhyx]{LadHyX, \'{E}cole polytechnique, CNRS, Institut Polytechnique de Paris, 91128 Palaiseau, France}
\address[onera]{DAAA, ONERA, Université Paris-Saclay, Meudon F-92190, France}

\begin{abstract}
The quasistatic approximation is a useful but questionable simplification for analyzing step instabilities during the growth/evaporation of vicinal surfaces. Using this approximation, we characterized in Part I of this work the effect on stability of different mechanisms and their interplay: elastic step-step interactions, the Schwoebel barrier, and the chemical coupling of the diffusion fields on adjacent terraces. In this second part, we present a stability analysis of the general problem without recourse to the quasistatic approximation. This analysis reveals the existence of a supplementary  mechanism, which we label the ``dynamics effect'' as it follows from accounting for all the convective and transient terms in the governing equations. This effect can be stabilizing or destabilizing depending on the ratio of step attachment/detachment kinetics to terrace diffusion kinetics. Further, we find that this dynamics effect remains significant in the slow deposition/evaporation regime, thereby invalidating the classical postulate underlying the quasistatic approximation. Finally, revisiting experiments of crystal growth on Si(111)-7\texttimes7 and GaAs(001), our analysis provides an alternative explanation of the observed step bunching, one that does not require the mechanisms previously invoked in the literature.
\end{abstract}

\end{frontmatter}

\newpage
\tableofcontents
\newpage

%% main text
%%%%%%%%%%%%%%%%%%%%%%%%%%%%%%%%%%%%%%%%%%%%%%%%%%%%%%%%%%%%%%%%
%%%%%%%%%%%%%%%%%%%%%%%%%%%%%%%%%%%%%%%%%%%%%%%%%%%%%%%%%%%%%%%%\
%%%%%%%%%%%%%%%%%%%%%%%%%%%%%%%%%%%%%%%%%%%%%%%%%%%%%%%%%%%%%%%%
%\linenumbers
\section{Introduction}
\label{ssec:quasistatic}

In the companion paper we formulated the problem of step bunching instability on vicinal surfaces and proceeded with the linear
stability analysis of its quasistatic approximation. This simplification permitted a mostly analytical treatment of the problem, thus
allowing a significant insight on the influence of the different mechanisms on stability. In the present work, we go beyond the  quasistatic approximation by solving stability of the general step-flow problem. This unveils a new stabilizing/destabilizing mechanism, which we refer to as the ``dynamics effect''.

As stated in Section~2  of Part I, the dimensionless formulation of the moving boundary problem for the
epitaxial crystal growth with straight steps, assuming small departures of the adatom density from its step equilibrium value $\rho^*_{eq}$
and a vanishingly small driving force $\digamma$ at each step, is given by the following system
\begin{empheq}[left=\empheqlbrace]{align}
\begin{aligned}
 \partial_t \rho_n &= \partial_{xx}\rho_n -\onu \rho_n + \oF ,\\ 
-\rho_n^- \dot{x}_{n+1} - (\partial_x\rho_n)^-&=\underbrace{\ok \left( \rho_n^- - 1  - \chi \Theta \llbracket \rho \rrbracket_{x_{n+1}} + \mathfrak{f}_{n+1} \right)}_{J_{n+1}^-} - \okp \llbracket \rho \rrbracket _{x_{n+1}},\\ 
\rho_n^+ \dot{x}_{n} + (\partial_x\rho_n)^+&=\underbrace{\ok S \left( \rho_n^+ - 1 - \chi \Theta \llbracket \rho \rrbracket_{x_{n}} + \mathfrak{f}_{n}\right)}_{J_{n}^+} + \okp \llbracket \rho \rrbracket_{x_{n}},\\
\dot{x}_n&=\Theta (J_{n}^++J_{n}^-), \\ 
\mathfrak{f}_n &=\sum_{\substack{r\in\{-R,...,R\} \\ r \neq 0 } } \left\{\frac{\ob}{x_{n+r}-x_n} - \frac{\oa}{(x_{n+r}-x_n)^3}\right\},
\end{aligned}
\label{eq:fbvp2}
\end{empheq}
where $\rho_n$, $x_n$, and $x_{n+1}$ denote the nondimensional adatom density and step positions, respectively and $\mathfrak{f}_n$ is 
the elastic contribution to the configurational force acting of the $n$th step.

As detailed in Section~2.5 of Part I, the \textit{quasistatic approximation} consists in neglecting the dynamics terms: $\partial_t \rho_n$ in \eqref{eq:fbvp2}$_1$,  $\rho_n^{-} \dot{x}_{n+1}$ in \eqref{eq:fbvp2}$_{2}$ and $\rho_n^{+} \dot{x}_{n}$ in \eqref{eq:fbvp2}$_{3}$, resulting in considerable simplification
of the stability analysis. To the best of the author's knowledge, no previous
work has explored the consequences of the inclusion of these terms on the stability of steps based on the general equations \eqref{eq:fbvp2}, and this
is the goal of the present article, Part II of our study.  

In terms of method, the challenge raised by the stability analysis of \eqref{eq:fbvp2} lies in that both the adatom density function and their domains of definition are subject to perturbations. To address that issue we use an Arbitrary Lagrangian-Eulerian (ALE) mapping of \eqref{eq:fbvp2}, which allows us to formulate the linear stability problem on a periodic system of infinite size, subsequently rewritten on a unit cell (one terrace) by using Floquet-wave decomposition. The stability problem hence takes the form of a generalized eigenvalue problem, solved numerically using a Chebyshev collocation method.

That general theory reveals dynamics as a new stabilizing/destabilizing mechanism, which we characterize by isolating it from the other classical mechanisms. We find that, both under deposition and evaporation, the effect of dynamics on stability depends on the value of $\ok$ (expressing the ratio of step attachment/detachment kinetics to terrace diffusion kinetics). When combined with the other mechanisms (elasticity, Schwoebel and chemical effects), dynamics significantly modifies the quasistatic stability diagrams. These new data call for reinterpreting some experimental results. In this regard, we show that the general stability analysis provides a possible explanation for understanding occurrences of bunching on Si(111)-7\texttimes7 and GaAs(001).

The rest of the article is organized as follows: The general stability analysis of \eqref{eq:fbvp2} is presented in Section~\ref{sec:stability2} and its results are given in Section~\ref{sub:lsa2results}. In Section~\ref{sec:exp}, we reinterpret some experiments showing bunching. Finally, we discuss  in Section~\ref{sec:discussion} the quasistatic approximation and how our analysis compares with previous works on dynamics. We summarize our main results in the conclusion, Section~\ref{sec:conclusion}.

%%%%%%%%%%%%%%%%%%%%%%%%%%%%%%%%%%%%%%%%%%%%%%%%%%%%%%%%%%%%%%%%
%%%%%%%%%%%%%%%%%%%%%%%%%%%%%%%%%%%%%%%%%%%%%%%%%%%%%%%%%%%%%%%%
\section{Stability analysis including dynamics terms} 
\label{sec:stability2}

In this section, we perform the linear stability analysis of the step-flow problem \eqref{eq:fbvp2}. Unlike its counterpart in the quasistatic approximation (see Section 3 of the companion paper), which furnishes closed-form expressions for the growth rate of a perturbation, the present analysis accounting for the dynamics effect yields stability results that are numerical. The fundamental steady-state solution is computed in Section~\ref{sub:sss2}. This is followed in Section~\ref{sub:lsa2} by the stability analysis, which involves three steps: an arbitrary Lagrangian-Eulerian mapping of \eqref{eq:fbvp2}, the linearization of the resulting equations about the steady-state solution, and a Floquet-wave analysis of the linearized equations. This procedure yields, for each wavelength of instability, a generalized eigenvalue problem whose numerical resolution furnishes the growth rate associated to the mode of instability.

%%%%%%%%%%%%%%%%%%%%%%%%%%%%%%%%%%%%%%%%
\subsection{Steady-state solution} 
\label{sub:sss2}

Like in Section 3.1 of part I of this work, the fundamental solution of \eqref{eq:fbvp2} corresponds to the propagation of equidistant steps with the $n$th step position given by  $\xz_n(t)=n+\vz t$ and where the adatom density $\rz(x,t)$ takes the form:
\begin{equation}
\rz(x,t)=\rtz\big(x-\xz_n(t)\big),
\end{equation}
for  $x \in(\xz_n(t),\xz_n(t)+1)$ with $\rz$ defined on $(0,1)$. After inserting these expressions in \eqref{eq:fbvp2}, one can obtain, by solving \eqref{eq:fbvp2}$_{1-3}$, an analytical expression of $\rtz$ which involves the unknown step velocity $\vz$. The subsequent insertion of the resulting currents $J_n^-$ and $J_n^+$ into \eqref{eq:fbvp2}$_{4}$ yields an equation for $\vz$. In the particular case of deposition only ($\onu=0$), this equation can be solved analytically and, as it turns out, one recovers the velocity of the quasistatic steady-state solution $\vz=\oF \Theta$. By contrast, in the general case ($\onu \neq 0$), this is a transcendental equation which must be solved numerically.

%%%%%%%%%%%%%%%%%%%%%%%%%%%%%%%%%%%%%%%%
\subsection{Linear stability} 
\label{sub:lsa2}

\subsubsection*{Arbitrary Lagrangian-Eulerian formulation}
We proceed to the linear stability of the steady-state solution of \eqref{eq:fbvp2}. To circumvent the difficulty of moving
boundaries, we use an Arbitrary Lagrangian-Eulerian (ALE) formulation by which we substitute the ALE variable  $u \in (0,1)$ to the spatial variable $x \in \big( x_n(t), x_{n+1}(t) \big)$  through the change of variable

\begin{equation} 
\label{eq:diffeo}
u:=\frac{x-x_n(t)}{x_{n+1}(t)-x_n(t)}.
\end{equation}
We then introduce the ALE adatom density $\rt_n(u,t)$ on terrace $n$th defined on $(0,1) \times \mathbb{R}^+$ by
\begin{equation} \label{eq:lagdens}
\rt_n(u,t):=\rho_n\big(u\big[x_{n+1}(t)-x_n(t)\big]+x_n(t),t\big).
\end{equation}
Making use of the relations between the partial derivatives of $\rho_n$ and of $\rt_n$,\footnote{
From \eqref{eq:diffeo}, \eqref{eq:lagdens}:
$\partial_t \rt_n  = \big( \dot{x}_n+(\dot{x}_{n+1}-\dot{x}_n)u \big) \partial_x \rho_n + \partial_t \rho_n , \;  
\partial_u \rt_n  = (1+x_{n+1}-x_n ) \partial_x \rho_n , \;   
\partial_{uu} \rt_n  = (1+x_{n+1}-x_n )^2 \partial_{xx} \rho_n. $}
\eqref{eq:fbvp2} is rewritten with $\rt_n$ as

\begin{empheq}[left=\empheqlbrace]{align} 
\begin{aligned}
 s_n^2 \partial_t \rt_n& = \partial_{uu}\rt_n +s_n \big(  \dot{x}_{n}+ (\dot{x}_{n+1}-\dot{x}_{n})u \big)  \partial_{u}\rt_n 
 + s_n^2 \big(-\onu \rt_n + \oF \big),\\ 
- s_n \rt_n^- \dot{x}_{n+1}  -(\partial_u\rt_n)^-&= s_n \Big[
\underbrace{\ok \Big( \rt_{n}^--1- \chi  \Theta \big( \rt_{n+1}^+-\rt_{n}^- \big) +f_{n+1}
 \Big)}_{\tilde{J}_{n+1}^-} - \okp \big(\rt_{n+1}^+-\rt_n^- \big)
 \Big] ,\\ 
  s_n \rt_n^+\dot{x}_{n} +(\partial_u\rt_n)^+&=s_n \Big[ 
\underbrace{\ok S \Big( \rt_{n}^+-1-\chi \Theta \big( \rt_n^+-\rt_{n-1}^- \big) +f_n
 \Big)}_{\tilde{J}_{n}^+} + \okp \big(\rt_{n}^+-\rt_{n-1}^- \big) \Big],\\ 
\dot{x}_n&=\Theta (\tilde{J}^+_{n} + \tilde{J}^-_{n}),
\end{aligned}
\label{eq:aftercov}
\end{empheq}
where $s_n(t):=x_{n+1}(t)-x_n(t)$ and the superscripts $(^+)$ and $(^-)$ denote evaluations  at $u=0$ and $u=1$, respectively: $\rt_{n}^+(t):=\rt_{n}(0,t)$, $\; \rt_{n}^-(t):=\rt_{n}(1,t)$, $\; (\partial_u\rt_n)^+(t):= \partial_u\rt_n(0,t)$ and $(\partial_u\rt_n)^-(t):=\partial_u\rt_n (1,t)$. 

\subsubsection*{Linear perturbation equations}
Noting that, for the steady-state solution, $u= x-\xz_n(t)$, the ALE form of the principal solution is 
$\rtz(u)$ with $\rtz$ obtained in Section~\ref{sub:sss2}. The linear perturbation equations are derived by considering the following perturbed state:
\begin{empheq}[left=\empheqlbrace]{align} \label{eq:lin}
\begin{aligned}
x_n(t)&=  n+ \vz t + \varepsilon \delta x_n(t)+o(\varepsilon), \\
\rt_n(u,t)&= \rtz(u)+ \varepsilon \delta \rt_n(u,t)+o(\varepsilon),
\end{aligned}
\end{empheq}
where $\varepsilon$ is a small parameter. Denoting by $\mathbf{q}_n(u,t):=\big(\delta x_n(t),\delta \rt_n(u,t)\big)$ the vector of the perturbed step positions and adatom densities, insertion of \eqref{eq:lin} in \eqref{eq:aftercov} and collection of terms of order $\varepsilon$ yields a linear system\footnote{Only nearest step interactions
are considered here, i.e., $R=1$ in the elastic configurational force $\mathfrak{f}_n$ in \eqref{eq:fbvp2}$_4$; the influence of $R>1$ was discussed
in the companion paper.} for $\mathbf{q}_n$,
\begin{equation}
\mathcal{A}\big(\mathbf{q}_{n-1},\mathbf{q}_n,\mathbf{q}_{n+1},\mathbf{q}_{n+2} \big) =
\mathcal{B} \big( \partial_t\mathbf{q}_n,\partial_t \mathbf{q}_{n+1} \big),
\label{eq:lpe}
\end{equation}
where and $\mathcal{A}$ and $\mathcal{B}$ are time-independent linear operators involving $u$-derivatives of $\delta \rt_n$, whose expressions are given in \ref{app:lsadetails}.

\subsubsection*{Floquet-wave analysis}
The linear differential equations in \eqref{eq:lpe} have a spatial variable $v=u+n \in \mathbb{R}$, where $u \in (0,1)$ is the local variable and $n \in \mathbb{Z}$. They are translationally invariant, since the linear operators $\mathcal{A}$ and $\mathcal{B}$ in \eqref{eq:lpe} are independent of
$n$, i.e., are invariant under $m$-terrace translations (for all $m  \in \mathbb{Z}$). This space periodicity of the differential operators allows us, according to {\it Floquet-wave theory}, to write the solutions of \eqref{eq:lpe} in the form: $\drt_n(u,t)=\delta \check{\rho}(u,t) \exp( ikn)$ and $\delta x_n(t) = \delta \check{x}(t) \exp (i k n)$ where the 
wavenumber $k \in [-\pi,\pi]$. In addition the system is autonomous, i.e., the linear operators $\mathcal{A}$ and $\mathcal{B}$ in \eqref{eq:lpe} are time-independent, which leads to solutions in the form: $\delta \check{\rho}(u,t)=\delta \hat{\rho}(u) \exp (\lambda t)$ and $\delta \check{x}(t) = \delta \hat{x} \exp (\lambda t)$, where $\lambda$ is the perturbation's growth rate. Combining these results, the solutions of \eqref{eq:lpe} can be written as follows:
\begin{equation}
\delta x_n(t) = \delta \hat{x} \exp (i k n + \lambda t),  \quad
 \delta \rt_n(u,t)  = \delta \hat{\rho}(u) \exp (i k n + \lambda t).
\label{eq:bw}
\end{equation}
Inserting \eqref{eq:bw} in \eqref{eq:lpe} yields a generalized eigenvalue problem associated to wavenumber $k$,
\begin{equation} \label{eq:lpebw}
\hat{\mathcal{A}}_k \hat{\mathbf{q}} = \lambda \hat{\mathcal{B}}_k \hat{\mathbf{q}},
\end{equation}
where $\hat{\mathbf{q}}(u):=\big(\delta \hat{x}, \delta \hat{\rho}(u)\big)$ and $\hat{\mathcal{A}}_k$ and $\hat{\mathcal{B}}_k$ are linear operators derived
from $\mathcal{A}$ and $\mathcal{B}$ whose corresponding expressions are furnished in \ref{app:lsadetails}.

As detailed in \ref{app:numerical}, the numerical resolution of \eqref{eq:lpebw} for a given $k$ furnishes a set of complex eigenvalues $\calE(k)$, from which we obtain the critical growth rate  $\mathrm{Re}\big(\lambda(k)\big)$ where $\lambda(k)$ is the eingenvalue in $\calE(k)$ with maximum real part. The train of equidistant steps is linearly stable with respect to bunching when $\mathrm{Re}\big(\lambda(k)\big)<0$ for all $k \in [-\pi,\pi]$ and unstable otherwise. Since $\lambda(-k)=\overline{\lambda(k)}$, $\mathrm{Re}(\lambda(k)\big)$ needs only to be studied on $[0, \pi]$, which establishes the dispersion relation of the problem at hand.

%%%%%%%%%%%%%%%%%%%%%%%%%%%%%%%%%%%%%
%%%%%%%%%%%%%%%%%%%%%%%%%%%%%%%%%%%%%
\section{Results}
\label{sub:lsa2results}

The stability analysis presented in Section~\ref{sec:stability2} allows, by contrast with its quasistatic counterpart in Part I, to capture the effect of dynamics on step bunching. Indeed, it turns out that the inclusion of the dynamics terms in the stability analysis reveals a new stabilizing/destabilizing mechanism, which we call \emph{the dynamics effect}.

 We first characterize, in Section~\ref{sub2:edt}, the specific effect of dynamics on stability by isolating it from the other mechanisms. In Section~\ref{sub2:dynCoeff} we find how the dynamics effect scales with the material and operational parameters $\Theta$, $\oF$, and $\onu$, which allows us to discuss its importance relative to the other mechanisms. Finally, in Section~\ref{sub:inconsistencies} we show how the accounts of dynamics modifies the stability predictions associated to the classical mechanisms.

\begin{figure*}[tb]
\begin{center}
\includegraphics[width=0.4\textwidth]{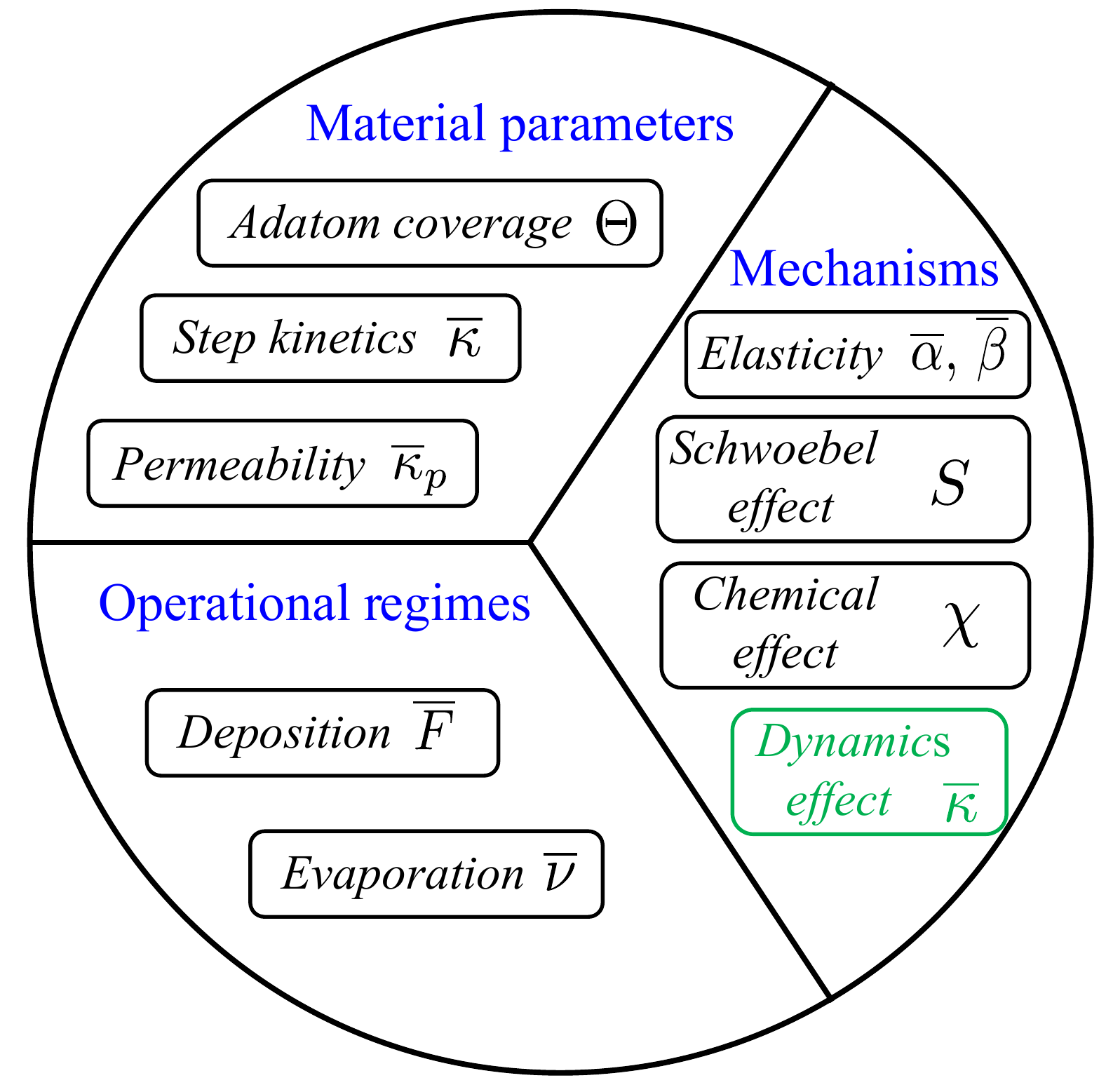}
\caption{Modification of the diagram in Figure~4 of Part I that shows the complete set of factors governing
the stability against bunching, including the dynamics effect as a stabilizing/destabilizing mechanism for different regimes of the step 
kinetics parameter $\ok$.}
\label{fig:mech2} 
\end{center} 
\end{figure*}

To highlight the changes brought by the present general theory, the complete set of factors that govern
the stability against bunching, including the dynamics effect is shown in Figure~\ref{fig:mech2}. In the present section, we do not consider any particular material and select generic values or ranges for the material and operational parameters. These values are in line with the estimations performed for materials like Si and GaAs detailed in \ref{app:parameters}.
%%%%%%%%%%%%%%%%%%%%%FIGURE 03%%%%%%%%%%%%%%%%%%%%%%%%%%%%%%%
%%%%%%%%%%%%%%%%%%%%%%%%%%%%%%%%%%%%%%%%%%%%%%%%%%%%%%%%%%
%%%%%%%%%%%%%%%%%%%%%%%%%%%%%%%%%%%%%%%%%%%%%%%%%%%%%%%%%%%%%%%%%%%%%%%%
%%%%%%%%%%%%%%%%%%%%%%%%%%%%%%%%%%%%%%%%%%%%%%%%%%%%%%%%%%%%%%%%%%%%%%%%

%%%%%%%%%%%%%%%%%%%%%%%%%%%%%%%%%%%%%
\subsection{Effect of the dynamics in both the deposition and evaporation regimes} 
\label{sub2:edt}

\begin{figure*}[tb]
\begin{center}
\includegraphics[width=0.8\textwidth]{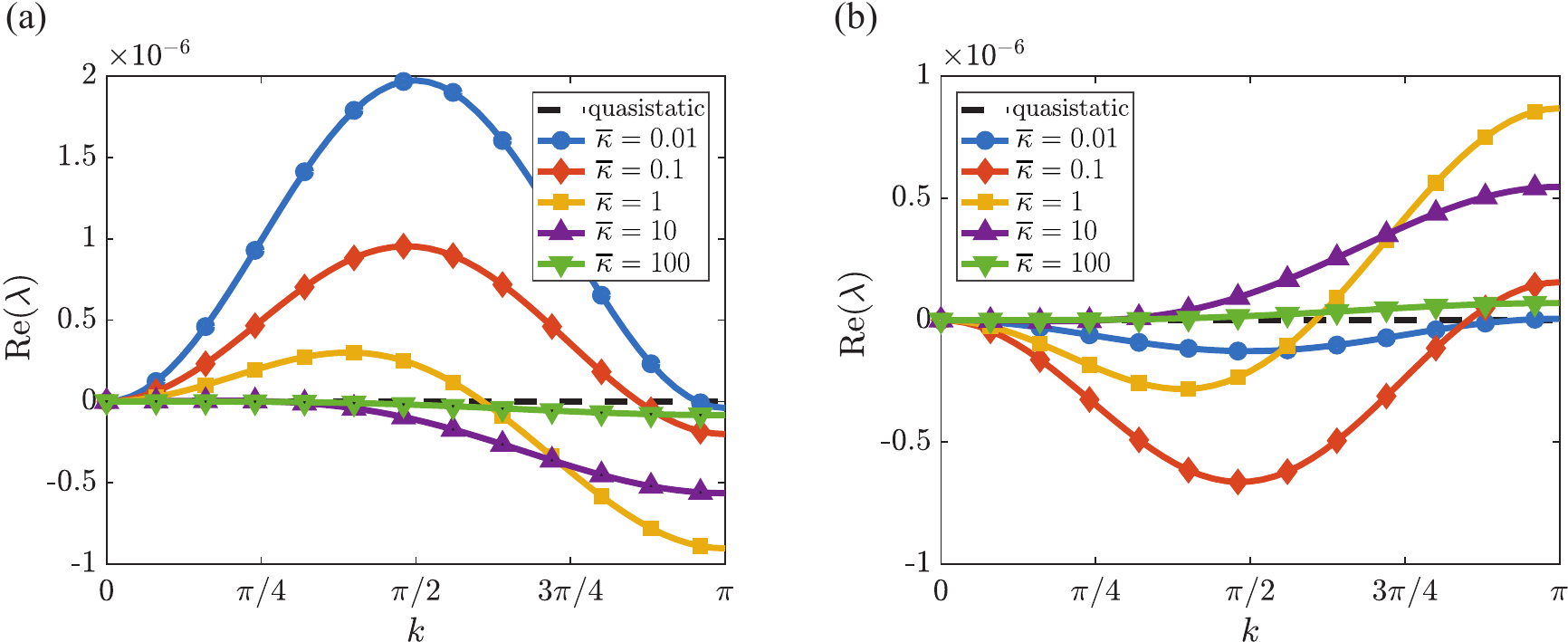} 
\caption{Dispersion curves: $\mathrm{max (Re}(\lambda))$ vs wavenumber $k$, showing the influence of the dynamics effect on stability in the absence of elasticity ($\oa = 0,\; \ob = 0$), chemical effect ($\chi = 0$), permeability ($\okp = 0$) and a Schwoebel factor $S=1$ under: (a) deposition ($\oF=0.01$, $\onu=0$) and (b) evaporation ($\onu=0.01$, $\oF=0$). Results are plotted for five different values of step kinetics $\ok$ and compared to the quasistatic case
where: $\mathrm{max (Re}(\lambda(k))=0,\ \forall k \in [0,\pi]$.}
\label{fig:dpokp} 
\end{center}
\end{figure*}

To analyze the effect  of dynamics on stability, we disable all the other stabilizing/destabilizing mechanisms such as elasticity ($\oa = 0$, $\ob = 0$), chemical effect ($\chi = 0$) and Schwoebel effect ($S=1$). Furthermore, steps are assumed impermeable ($\okp = 0$). The effect of dynamics on the different modes of bunching is given by the dispersion curves: $\mathrm{Re}\big(\lambda(k)\big)$ as a function of the wavenumber $k$.  As can be seen on Figure~\ref{fig:dpokp},%
%%%%%%%%%%%%%%%%%%%%%FIGURE 04%%%%%%%%%%%%%%%%%%%%%%%%%%%%%%%
%%%%%%%%%%%%%%%%%%%%%%%%%%%%%%%%%%%%%%%%%%%%%%%%%%%%%%%%%%
%%%%%%%%%%%%%%%%%%%%%%%%%%%%%%%%%%%%%%%%%%%%%%%%%%%%%%%%%%
%%%%%%%%%%%%%%%%%%%%%%%%%%%%%%%%%%%%%%%%%%%%%%%%%%%%%%%%%%
 the step kinetics $\ok$ plays a determinant role in the stabilizing/destabilizing influence of dynamics. 

Under deposition (see Figure~\ref{fig:dpokp}(a)), dynamics has a  stabilizing effect for high values of $\ok$ ($\ok \gg 1$, tends to the diffusion-limited regime\footnote{%
Recall from the companion paper that the notions of attachment/detachment (a/d)-limited regime and diffusion-limited regime refer to the two kinetic processes: a/d at steps and diffusion on terraces. $\ok:=\kappa_- L_0/D$ can be seen as the ratio of a characteristic step a/d velocity $\kappa_-$ to a characteristic diffusion velocity $D/L_0$. Then, $\ok \ll 1$ corresponds to situations where a/d is the limiting kinetic process and $\ok \gg 1$ is associated to cases the terrace diffusion is limiting. 
})  and a destabilizing effect for low values of $\ok$ ($\ok \ll 1$, tends to the attachment/detachment-limited regime). Interestingly, in the latter the most unstable modes are those of intermediate wavelength ($k \sim \pi/2$). This point is in stark contrast with the characteristics of the other destabilizing mechanisms, that predominantly affect step-pairing (see the discussion in Section~4 of the companion paper) and this is the reason for a special behavior in the maturing phase of the bunches when those are caused by the dynamics instability \citep{Benoit2020}.
Conversely, the effect of step kinetics is reversed for evaporation (see Figure~\ref{fig:dpokp}(b)), where low values of $\ok$ (i.e., $\ok \ll 1$) correspond to a stabilizing effect while intermediate and high values of $\ok$  (i.e., $\ok \gtrsim 1$) are destabilizing.
These results are summarized in Table~\ref{tab:dynInfl}, which completes Table~1 of the companion paper.
%%%%%%%%%%%%%%%%%%%%%%%%%%%%%%%%TABLE 01%%%%%%%%%%%%%%%%%%%%%%%%%%%%%%%%%%
\begin{table*}[tb]
\begin{center}
\begin{tabular}{l c c c c c c c}
\hline
& ES & iES & CE & DDE & MME & {\color{blue} DYN}  & {\color{blue} DYN} \\
& $S>1$ & $S<1$ & $\chi$ & $\oa$ & $\ob$ & $\ok \ll 1$ & $\ok \gg 1$ \\
\hline
\hline
Deposition & $\mathcal{S}$ & $\mathcal{D}$ & $\mathcal{D}$ & $\mathcal{S}$ & $\mathcal{D}$ & $\mathcal{D}$ & $\mathcal{S}$\\
\hline
Evaporation &  $\mathcal{D}$ & $\mathcal{S}$ &$\mathcal{S}$  & $\mathcal{S}$ & $\mathcal{D}$ & $\mathcal{S}$ & $\mathcal{D}$\\
\hline
\end{tabular}
\caption{Effects of each of the basic mechanisms, including dynamics, on the onset of the bunching instability.  ES refers to the Ehrlich--Schwoebel effect, iES to its inverse, CE to the chemical effect, DDE to dipole-dipole elastic interactions, MME to their monopole-monopole counterparts
and {\color{blue} DYN} to the influence of dynamics.  $\mathcal{S}$ stands for stabilizing and $\mathcal{D}$ for destabilizing.}
\label{tab:dynInfl}
\end{center}
\end{table*}
%%%%%%%%%%%%%%%%%%%%%%%%%%%%%%%%%%%%%%%%%%%%%%%%%%%%%%%%%%%%%%%%%%%%%%%%
%%%%%%%%%%%%%%%%%%%%%%%%%%%%%%%%%%%%%%%%%%%%%%%%%%%%%%%%%%%%%%%%%%%%%%%%

%%%%%%%%%%%%%%%%%%%%%%%%%%%%%%%%%%%%%%%%%%%%%%%%%%%%%%%%%%%%%%%%%%%%%%%%
\subsection{Scalings related to dynamics} 
\label{sub2:dynCoeff}

\begin{figure*}[h!]
\begin{center}
\includegraphics[width=0.8\textwidth]{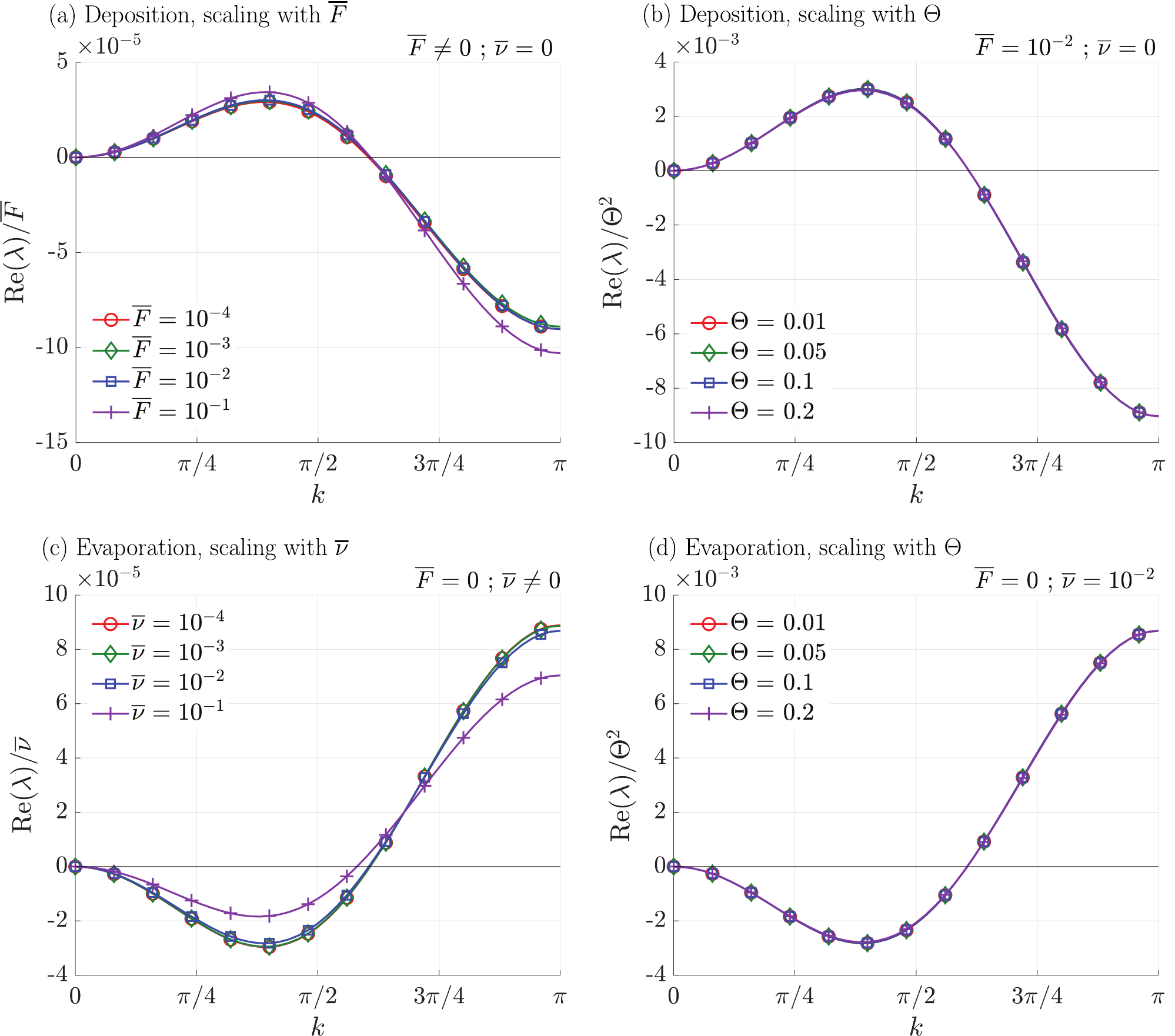}
\caption{ Scaling of the dispersion curves $\mathrm{Re}\big(\lambda(k)\big)$ with: (a) deposition rate $\oF$, (c) 
evaporation rate $\onu$ and (b), (d)
adatom coverage $\Theta$. Results, calculated for $\ok=1$, show the influence of the dynamics on stability in the absence of elasticity ($\oa = 0,\; \ob = 0$), chemical effect ($\chi = 0$), permeability ($\okp = 0$) and with a Schwoebel factor $S=1$. In (a) and (c), $\Theta=0.01$ }
\label{fig:dcScaling}
\end{center}
\end{figure*}

A parametric study with $\oF$, $\onu$ and $\Theta$ of the dispersion curves related to the dynamics effect reveals that these parameters do not alter the shape of the dispersion curve, but simply act as scaling factors for $\mathrm{Re} (\lambda)$. As can be seen on Figure~\ref{fig:dcScaling},
%%%%%%%%%%%%%%%%%%%%%FIGURE 05%%%%%%%%%%%%%%%%%%%%%%%%%%%%%%%
%%%%%%%%%%%%%%%%%%%%%%%%%%%%%%%%%%%%%%%%%%%%%%%%%%%%%%%%%%
%%%%%%%%%%%%%%%%%%%%%%%%%%%%%%%%%%%%%%%%%%%%%%%%%%%%%%%%%%
%%%%%%%%%%%%%%%%%%%%%%%%%%%%%%%%%%%%%%%%%%%%%%%%%%%%%%%%%%
showing the dispersion curves for different values of  $\oF$ and $\Theta$ under deposition and  $\onu$ and $\Theta$ under evaporation, the growth rate scales linearly with the  deposition/evaporation rates ($\mathrm{Re} (\lambda) \propto \oF$ and $ \mathrm{Re} (\lambda) \propto \nu$) while it scales quadratically with the adatom coverage ($\mathrm{Re} (\lambda) \propto \Theta ^2$).

The scalings of the dynamics contribution to stability with $\oF$, $\onu$ and $\Theta$ are summarized in Table~\ref{tab:coeff},%
%%%%%%%%%%%%%%%%%%%%%%%%%%%%TABLE 02%%%%%%%%%%%%%%%%%%%%%%%%%
%%%%%%%%%%%%%%%%%%%%%%%%%%%%%%%%%%%%%%%%%%%%%%%%%%%%%%%%%%
\begin{table*}[tb]
\begin{center}
\begin{tabular}{c c c c c c}
\hline
&   ES/iES & CE &  E  & {\color{blue} DYN} \\
& $S$ & $\chi$ & $\oa, \, \ob$ & $\ok$ \\ 
 \hline
 \hline
 $a$ (for $\oF, \, \onu$) & 1 & 1 & 0 & 1 \\
 \hline
 $b$  (for $\Theta$) & 1 & 2 & 1 & 2 \\
 \hline
\end{tabular}
\end{center}
\caption{The scalings of the contributions of the different mechanisms to the perturbation growth rate with the operating parameters $\oF$ and $\onu$ and the adatom coverage $\Theta$. ES/iES refers to the Ehrlich--Schwoebel barrier or its inverse, CE
to the chemical effect, E to the elastic step-step interactions, and {\color{blue} DYN} to the influence of dynamics. The contribution of each mechanism scales as $\re(\lambda) \propto \oF^a \Theta^b$ under deposition and $\re(\lambda) \propto \onu^a \Theta^b$ under sublimation, with the exponents $a$ and $b$ given in the table.}
\label{tab:coeff}
\end{table*}
%%%%%%%%%%%%%%%%%%%%%%%%%%%%%%%%%%%%%%%%%%%%%%%%%%%%%%%%%%%%
%%%%%%%%%%%%%%%%%%%%%%%%%%%%%%%%%%%%%%%%%%%%%%%%%%%%%%%%%%%%
 thus completing Table~2 of the companion paper.

%%%%%%%%%%%%%%%%%%%%%%%%%%%%%%%%%%%%%%%%%%%%%%%%%%%%%%%%%%%
\subsection{Modification of the quasistatic approximation stability results}
\label{sub:inconsistencies}

In this section, we examine successively the modifications caused by the dynamics effect on the stability results when: i) elasticity is included 
(dipole-dipole interactions),  ii) Schwoebel effect is included and iii) the chemical effect is accounted for.
 To this end, we compare the stability diagrams showing the domains of stable step propagation vs. step bunching as a function of the material and operational parameters. 

To be more accurate in the predictions of bunching, we introduce the concept of \emph{significant instability} corresponding to the regions where the growth rate $ \lambda_{max}:=\max_{k \in [0, \pi]} \mathrm{Re}\big(\lambda(k)\big)$ is sufficiently large for the bunching instability to develop within the deposition/evaporation of a thousand monolayers (i.e., a few hundred nanometers). Indeed, there are cases where $ \lambda_{max}$ is positive---thereby \textit{a priori} indicating unstable step propagation---but whose value is so small that the step bunches develop only after deposition/evaporation of a number of monolayers beyond that typically observed in experiments. We define the regime of \emph{significant instability} with the condition $\lambda_{max} \tau  \geq 1$, where $\tau$ is the dimensionless time associated to the deposition (resp. evaporation) of a thousand monolayers, i.e., $\tau_{dep}=1000/\oF \Theta$ (resp. $\tau_{eva}=1000/\onu \Theta$)\footnote{
Indeed, under deposition only ($\onu=0$), the deposition time  $T$ for one monolayer  is given by the equality, for an arbitrary surface area $A$, between that surface area and the one covered by the flux $F$ of adatoms during the time $T$, i.e., $A=F a^2 T A$. This furnishes  $T=1/Fa^2$ and thereby the dimensionless  deposition time for one monolayer is $1/\oF \Theta$. Similarly, under evaporation ($\oF=0$), one can easily show that the dimensionless time associated to the evaporation of one monolayer is  $1/\onu \Theta$.
}.

To conclude this section, we discuss the validity of the quasistatic approximation in the regimes of slow deposition $\oF \Theta \ll 1$ or evaporation 
$\onu \Theta \ll 1$, where the quasistatic approximation is classically---and as we will see unduly---invoked in the literature \citep[e.g.,][]{Krug2005,Michely2012}. 

\subsubsection*{Elasticity (dipole-dipole interactions)}

\begin{figure*}[tb]
\begin{center}
\includegraphics[width=0.9\textwidth]{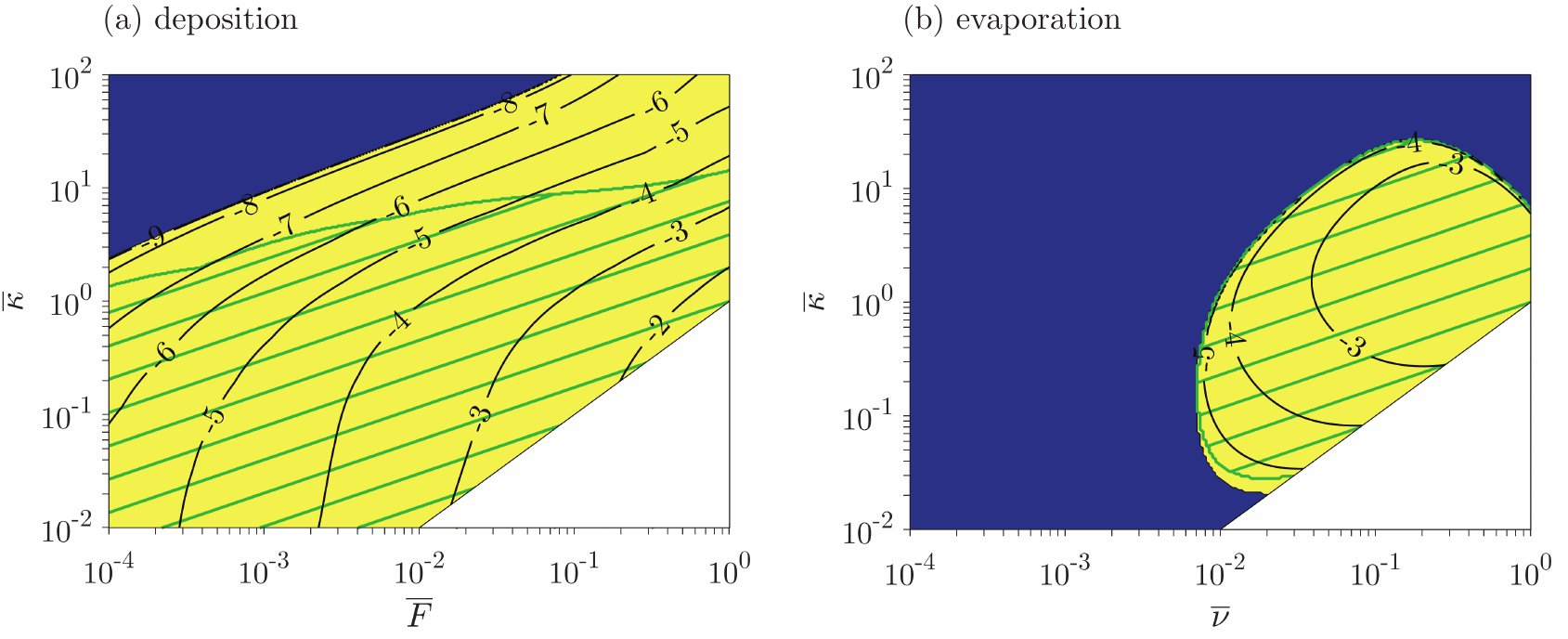}
\end{center}
\caption{Stability diagrams with respect to step bunching in the presence of dynamics effect and (dipole-dipole) elastic interactions under 
(a) deposition ($ \onu=0)$ 
and (b) evaporation ($\oF = 0)$. For both cases, $\Theta=0.2$, $\oa=10^{-4}$, and $\ob=0$, while permeability, Schwoebel and chemical effects are ignored ($\okp=0$, $S=1$, and $ \chi =0$). Blue, yellow and green hatched domains correspond to the \emph{stable}, \emph{unstable} and \emph{significantly unstable} regions, respectively. 
Recall for comparison that with the classical quasistatic model (only elastic interactions, no dynamics terms) 
step flow is stable (blue domain) everywhere. The white areas correspond to regions where growth/evaporation are not expected to take place in the step-flow regime. }
\label{fig:elasInfl} 
\end{figure*}

We first consider, in Figure~\ref{fig:elasInfl},%
%%%%%%%%%%%%%%%%%%%%%FIGURE 06%%%%%%%%%%%%%%%%%%%%%%%%%%%%%%%
%%%%%%%%%%%%%%%%%%%%%%%%%%%%%%%%%%%%%%%%%%%%%%%%%%%%%%%%%%
%%%%%%%%%%%%%%%%%%%%%%%%%%%%%%%%%%%%%%%%%%%%%%%%%%%%%%%%%%
%%%%%%%%%%%%%%%%%%%%%%%%%%%%%%%%%%%%%%%%%%%%%%%%%%%%%%%%%%
  the influence of the dynamics terms on the stability of steps interacting through dipole-dipole elastic interactions (homoepitaxy case: $\oa=10^{-4} and \ob=0$), while permeability, Schwoebel, and chemical effects are ignored ($\okp=0$, $S=1$, and $\chi =0$). 
Blue and yellow domains correspond to the stable ($ \lambda_{max}<0$) and unstable ($\lambda_{max} >0 $) regions, respectively. In the unstable domain, isolines display $ \log_{10}(\lambda_{max})$, indicating the magnitude of the most critical growth rate and the hatched green region shows the domain of significant instability as defined earlier in this section. The white area corresponds to combinations of $(\ok,\oF)$ and $(\ok,\onu)$  that are not within the hypothesis of validity of \eqref{eq:fbvp2} as they lead to adatom densities on the terraces that significantly depart from their equilibrium values (see Section~2.1 of the companion paper). In practice, we expect a breakdown of the step-flow regime in this region which makes the question of stability irrelevant. For comparison of these stability diagrams with the quasistatic case, recall that in the latter, because of the stabilizing effect of elasticity, the step propagation is stable both under deposition and evaporation for all parameter values.

Figure~\ref{fig:elasInfl}(a) shows the stability diagram, under deposition, in $(\oF, \ok)$ space, when dynamics terms are included.   
With $\oa=10^{-4}$ (the order of magnitude of interaction strength for a terrace width of $20~\mathrm{nm}$, see \ref{app:parameters}) we can see a region 
of instability, which is a first manifestation of the breakdown of the quasistatic approximation. Indeed, although we are in the regime $\oF \Theta \ll 1$, the dynamics terms have a non-negligible effect on stability.

Figure~\ref{fig:elasInfl}(b) shows the stability diagram, under evaporation, in the $(\onu, \ok)$ space, when dynamics terms are included, 
calculated for the same parameters as Figure~\ref{fig:elasInfl}(a) (i.e., $\oa=10^{-4}, \ob=0, S=1$, $\okp=0$, $ \chi =0$, and  $\Theta=0.2$).  
Again, because of the effect of dynamics, an unstable region appears for $\onu > 10^{-2}$ and $\ok$ in the intermediate range between 0.1 and 10. This region can be understood in the light of the effect of dynamics under evaporation shown on Figure~\ref{fig:dpokp}(b).

\subsubsection*{Schwoebel effect}

Next, we consider the interplay of the dynamics terms with the  Schwoebel effect (other mechanisms disabled : $\oa=0$, $\ob=0$, $\chi=0$, $\okp=0$). 

\begin{figure*}[tb]
\begin{center}
\includegraphics[width=0.9\textwidth]{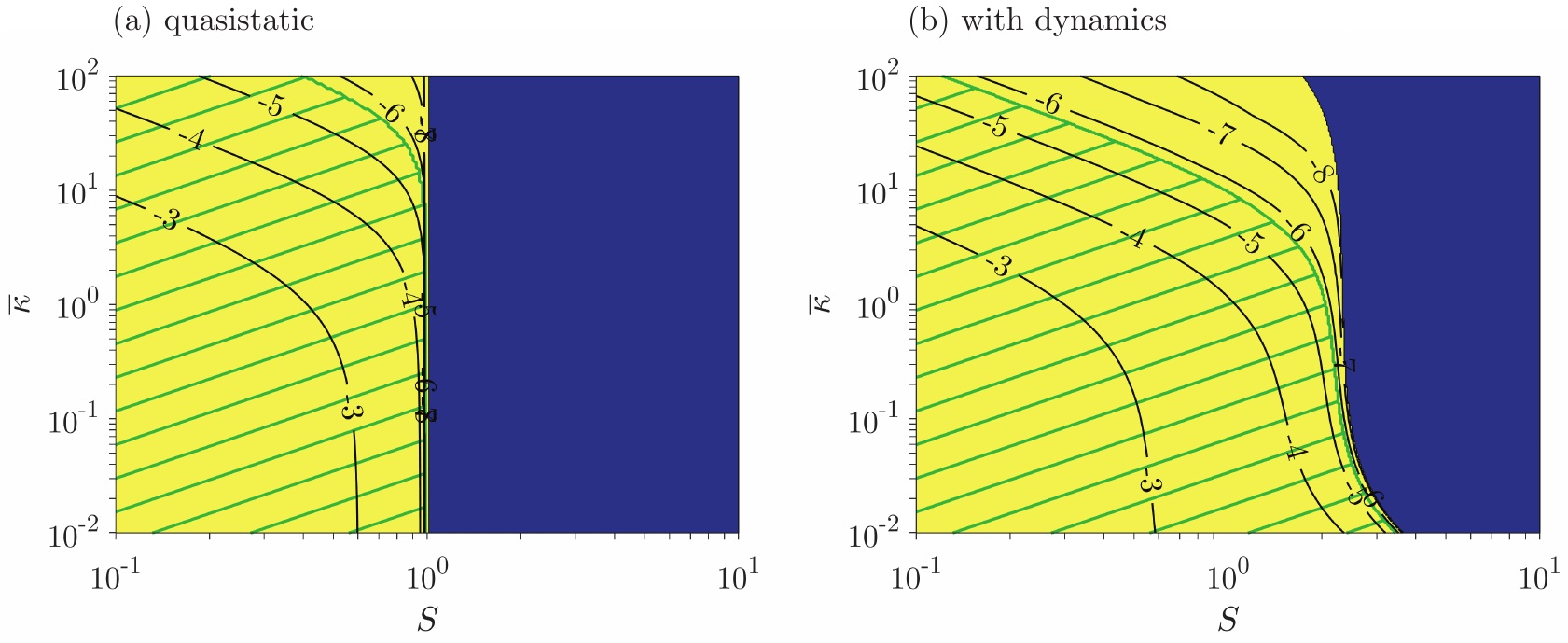}
\end{center}
\caption{Effect of the dynamics on the stability diagram of the Schwoebel effect under deposition. (a) Schwoebel effect only (b) Schwoebel effect with dynamics. In both cases $\oF=10^{-2}$ and $\Theta=0.2$; other mechanisms are disabled: $\oa=0$, $\ob=0$, $\okp=0$, $\chi=0$. (see Figure~\ref{fig:elasInfl} for the legends of colors and isolines) }
\label{fig:SchwobInflF} 
\end{figure*}

With Figure~\ref{fig:SchwobInflF}(a),%
%%%%%%%%%%%%%%%%%%%%%FIGURE 07%%%%%%%%%%%%%%%%%%%%%%%%%%%%%%%
%%%%%%%%%%%%%%%%%%%%%%%%%%%%%%%%%%%%%%%%%%%%%%%%%%%%%%%%%%
%%%%%%%%%%%%%%%%%%%%%%%%%%%%%%%%%%%%%%%%%%%%%%%%%%%%%%%%%%
%%%%%%%%%%%%%%%%%%%%%%%%%%%%%%%%%%%%%%%%%%%%%%%%%%%%%%%%%%
we recall the stability diagram associated with the Schwoebel effect under the quasistatic approximation for the deposition case (stable growth for $S>1$ and unstable one for $S<1$, see e.g., Table~\ref{tab:dynInfl}). As can be seen on Figure~\ref{fig:SchwobInflF}(b), the effect of dynamics is to extend the significantly unstable region beyond $S=1$ for slow step kinetics ($\ok<1$) and reduce it below $S=1$ for fast step kinetics ($\ok>10$). While this modification is consistent with the effect of dynamics shown on Figure~\ref{fig:dpokp}(a), it is important to note that the stability of the Schwoebel effect is substantially modified by dynamics in the very regime of small deposition rate (here $\oF \Theta =2 \times 10^{-3} \ll 1$).

\begin{figure*}[tb]
\begin{center}
\includegraphics[width=0.9\textwidth]{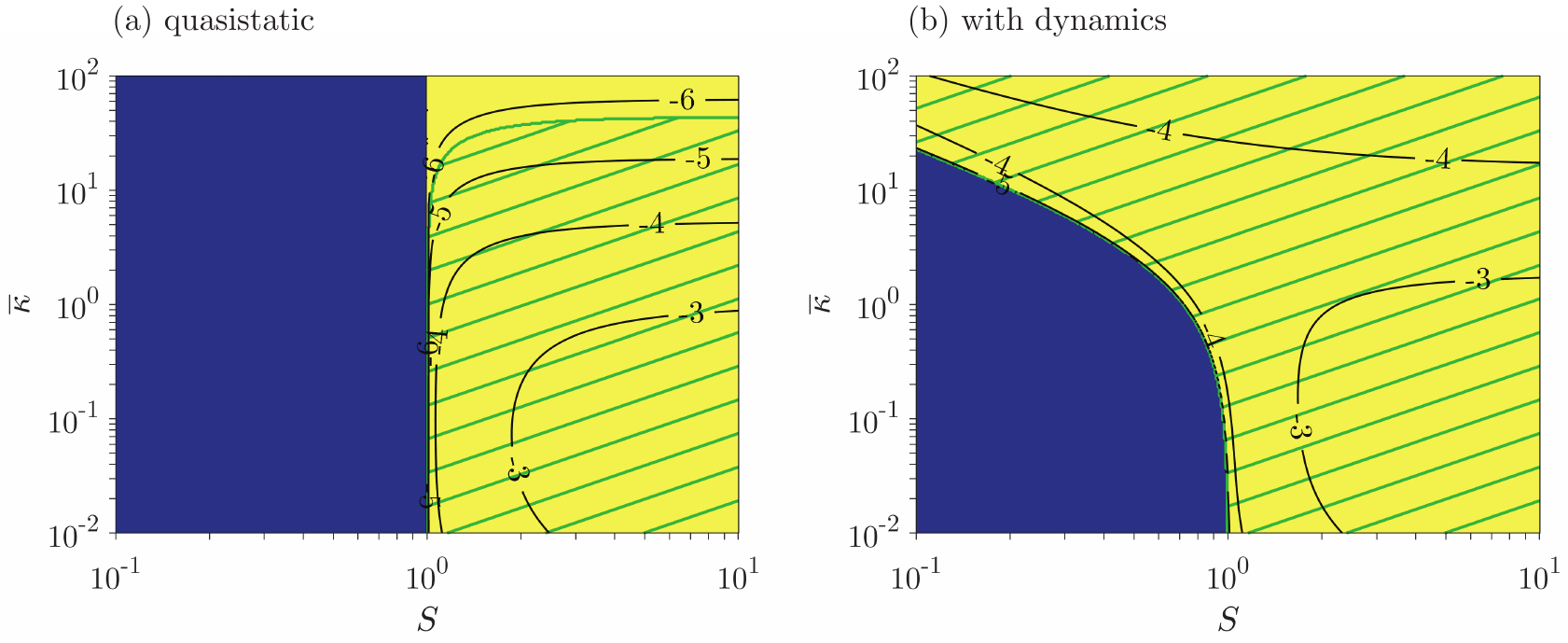}
\end{center}
\caption{Effect of the dynamics on the stability diagram of the Schwoebel effect under evaporation. (a) Schwoebel effect only. (b) Schwoebel effect with dynamics. In both cases $\onu=10^{-2}$ and $\Theta=0.2$; other mechanisms are disabled: $\oa=0$, $\ob=0$, $\okp=0$, $\chi=0$. (see Figure~\ref{fig:elasInfl} for the legends of colors and isolines) }
\label{fig:SchwobEvap} 
\end{figure*}

Similarly, under evaporation (see Figure~\ref{fig:SchwobEvap}) dynamics modifies the stability diagram of the Schwoebel effect with an extension of the unstable region in the domain $S<1$ for fast step kinetics ($\ok>1$).

%%%%%%%%%%%%%%%%%%%%%FIGURE 08%%%%%%%%%%%%%%%%%%%%%%%%%%%%%%%
%%%%%%%%%%%%%%%%%%%%%%%%%%%%%%%%%%%%%%%%%%%%%%%%%%%%%%%%%%
%%%%%%%%%%%%%%%%%%%%%%%%%%%%%%%%%%%%%%%%%%%%%%%%%%%%%%%%%%
%%%%%%%%%%%%%%%%%%%%%%%%%%%%%%%%%%%%%%%%%%%%%%%%%%%%%%%%%%

\subsubsection*{Chemical effect}

\begin{figure*}[tb]
\begin{center}
\includegraphics[width=0.9\textwidth]{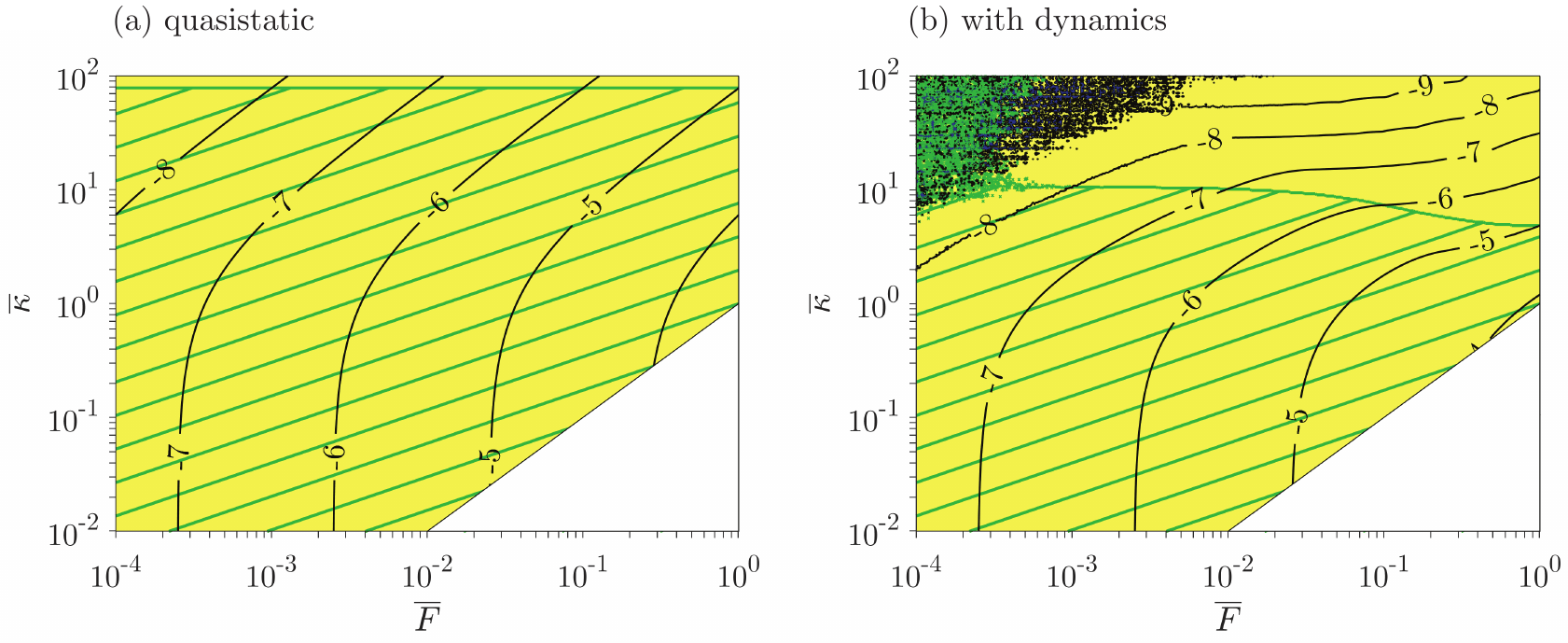}
\end{center}
\caption{Effect of the dynamics on the stability diagram of the chemical effect under deposition. (a) chemical effect only (b) chemical effect with dynamics. In both cases $\Theta=0.01$ and other mechanisms are disabled: $S=1$, $\oa=0$, $\ob=0$, and $\okp=0$. (see Figure~\ref{fig:elasInfl} for the legends of colors and isolines) }
\label{fig:ChemicalDepo} 
\end{figure*}

The chemical effect, which couples the diffusion fields on all terraces yields, within the quasistatic approximation, an unstable step flow under deposition and a stable one under evaporation (see Table~\ref{tab:dynInfl} and Section~4.2 of the companion paper). Figure~\ref{fig:ChemicalDepo}%
%%%%%%%%%%%%%%%%%%%%%FIGURE 09%%%%%%%%%%%%%%%%%%%%%%%%%%%%%%%
%%%%%%%%%%%%%%%%%%%%%%%%%%%%%%%%%%%%%%%%%%%%%%%%%%%%%%%%%%
%%%%%%%%%%%%%%%%%%%%%%%%%%%%%%%%%%%%%%%%%%%%%%%%%%%%%%%%%%
%%%%%%%%%%%%%%%%%%%%%%%%%%%%%%%%%%%%%%%%%%%%%%%%%%%%%%%%%%
shows that under deposition dynamics reduces the domain of significant instability where $\ok \gg 1$, which is consistent with its stabilizing effect in this regime (see Figure~\ref{fig:dpokp}(a)). Note nevertheless that no stable domain (in the sense $\lambda_{max}<0$)  appears because of the marginal instability of dynamics for long wavelength modes $k \rightarrow 0$ when $\ok \gg 1$. Under evaporation, dynamics has a destabilizing effect for $\ok \gg 1$, which however is not strong enough to reverse the stability of the chemical effect. As a result, both under the quasistatic approximation and with dynamics, the stability diagram associated to the chemical effect under evaporation shows stable step-flow growth for all parameter values.

\subsubsection*{Relative importance of dynamics for different values of $\oF$ and $\Theta$}

The above discussion	on the interplay between the dynamics effect and the three fundamental mechanisms of step-flow growth (see Figure~\ref{fig:mech2}) 
shows that even in the regimes of slow deposition or slow evaporation (defined by $\oF \Theta \ll 1$ and $\onu \Theta \ll 1$, respectively), where the quasistatic approximation is classically invoked in the literature, the effect of the dynamics terms is far from negligible. While this already establishes the failure of the quasistatic approximation, we can still wonder whether the effect of dynamics becomes negligible as $\oF \rightarrow 0$ or  $\onu \rightarrow 0$. It turns out not being the case in general, which confirms that, while the quasistatic approximation has been based on the postulate that it is valid for slow deposition/evaporation, the latter is fundamentally erroneous. To discuss the relative importance of dynamics for vanishingly small deposition/evaporation rates, we go back to the scaling of the different mechanisms with $\oF$, $\onu$ and $\Theta$ given in Table~\ref{tab:coeff}. 

Considering first elasticity vs. dynamics, we note that the effect of elasticity is independent of the deposition/evaporation rates while that of the dynamics is linear in that rate. As a result, the latter becomes negligible compared to the former as $\oF,\onu \rightarrow 0$ (see Figure~\ref{fig:elasInfl}). Note however that this is not specific to dynamics as the same conclusion is drawn when comparing the Schwoebel or chemical effects with elasticity. This scaling difference is the basis for the distinction between the \emph{energetic mechanisms}, qualifying elasticity, and the \emph{kinetic mechanisms}, applicable to the Schwoebel, chemical and dynamics effects.

\begin{figure*}[tb]
\begin{center}
\includegraphics[width=0.9\textwidth]{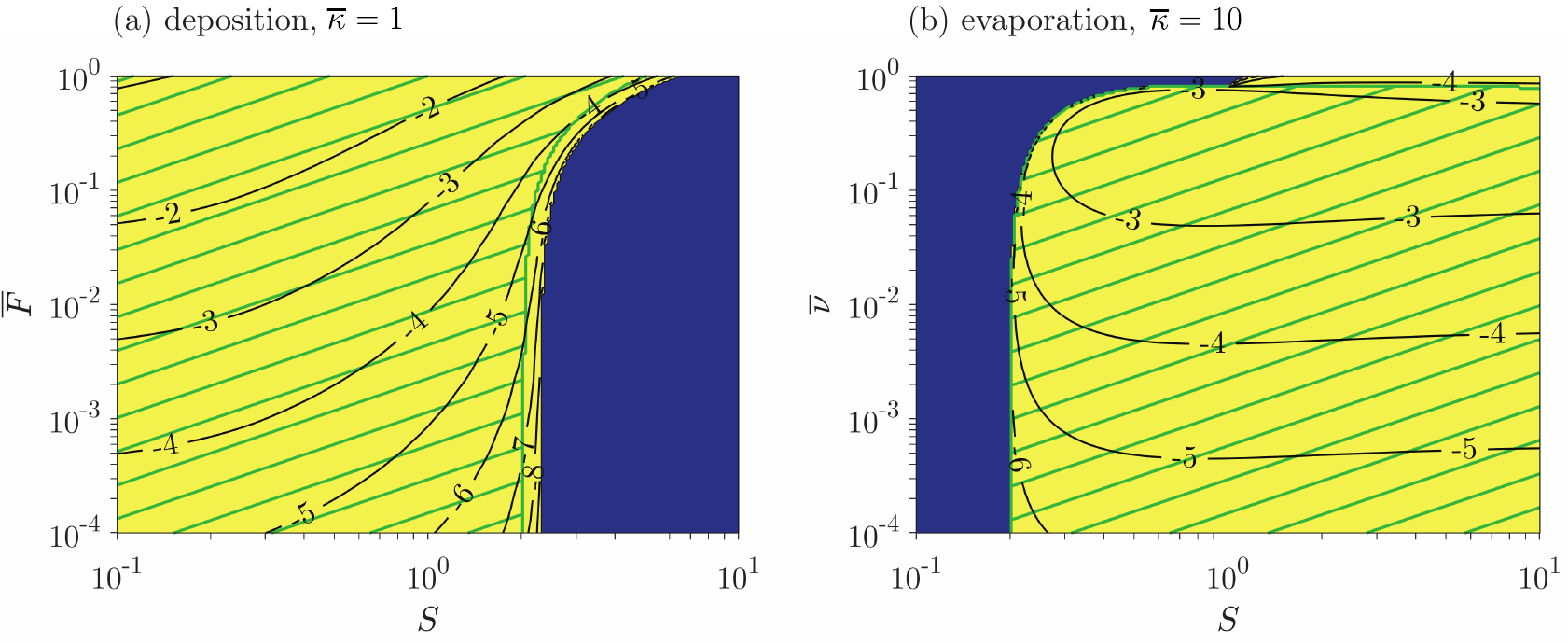}
\end{center}
\caption{Influence of the depositon/evaporation rates on the stability diagram combining Schwoebel and dynamics effects. (a) under deposition (with $\ok=1$) (b) under evaporation (with $\ok=10$). In both cases $\Theta=0.01$ and other mechanisms are disabled: $\oa=0$, $\ob=0$, $\chi=0$ and $\okp=0$. (see Figure~\ref{fig:elasInfl} for the legends of colors and isolines) }
\label{fig:SchwoebelRate} 
\end{figure*}

Because the three kinetic mechanisms have the same scaling with $\oF$ and $\onu$, their relative importance remains unchanged as $\oF,\onu \rightarrow 0$. As a numerical confirmation, we have observed that the stability diagrams of Figures~\ref{fig:SchwobInflF} and \ref{fig:SchwobEvap} remain unchanged for various $\oF$, $\onu$ across multiple decades below $\oF,\onu=10^{-1}$. To illustrate that point, we can see on Figure~\ref{fig:SchwoebelRate}%
%%%%%%%%%%%%%%%%%%%%%FIGURE 09%%%%%%%%%%%%%%%%%%%%%%%%%%%%%%%
%%%%%%%%%%%%%%%%%%%%%%%%%%%%%%%%%%%%%%%%%%%%%%%%%%%%%%%%%%
%%%%%%%%%%%%%%%%%%%%%%%%%%%%%%%%%%%%%%%%%%%%%%%%%%%%%%%%%%
%%%%%%%%%%%%%%%%%%%%%%%%%%%%%%%%%%%%%%%%%%%%%%%%%%%%%%%%%%
how the critical Schwoebel barrier (i.e., $S$ at which stability is reversed) varies with $\oF$ under deposition and $\onu$ under evaporation. That this critical barrier is independent of $\oF$ and $\onu$ when these are below $0.1$ shows that the effect of dynamics (relative to Schwoebel) remains significant for $\oF,\onu \rightarrow 0$. The same conclusion can be drawn from Figure~\ref{fig:ChemicalDepo} when comparing the relative importance of dynamics on the chemical effect. With these elements, it is clear that the conditions $\oF \Theta \ll 1$ and $\onu \Theta \ll 1$ or even the more restrictive ones $\oF \Theta \rightarrow 0$ and $\onu \Theta \rightarrow 0 $ are not sufficient for considering the effect of dynamics negligible. Whether dynamics contributes significantly or not to the (in)stability of the step flow for a particular experiment, depends, like for any other mechanism, on the whole set of material and operational parameters.

\subsubsection*{Effect of dynamics on the mode of instability}

The above comparison between the stability diagrams under the quasistatic approximation and their counterpart with dynamics reveals domains where stability changes, i.e., where the sign of $\lambda_{max}$ is reversed (or where $\lambda_{max}$ crosses the critical value at which step flow is or not significantly unstable). Whereas that aggregate information is sufficient to distinguish between stable step-flow growth and step bunching, the initial mode of instability (e.g., step pairing vs. long wavelength bunching) depends on the whole dispersion curve $\mathrm{Re}\big(\lambda(k)\big)$. In \citet{Guin2018}, we further discuss how dynamics affects these dispersion curves on some particular examples. This reveals that even in domains where dynamics does not reverse stability, it may significantly affect the shape of the underlying dispersion curves, i.e., how bunches grow with time.

%%%%%%%%%%%%%%%%%%%%%%%%%%%%%%%%%%%%%%%%%%%%%%%%%%%%%%%%%%%%%%%%
\section{Reinterpreting some experiments}
\label{sec:exp}

In this section, we consider two specific materials where step bunching is observed under deposition and discuss how the account of dynamics and chemical effects may provide an explanation to bunching alternative to those proposed in the literature. To that end, we consider all the mechanisms together, whose strength we estimate through an extended research of the values of the parameters at play in the experiments. The reader may refer to \ref{app:parameters} for the estimation of the material parameters.

The experiments considered are the epitaxial growth of GaAs(001) at  600$^{\circ}\mathrm{C}$-700$^{\circ}\mathrm{C}$ \citep{Hata1993,Kasu1992,Ishizaki1994,Pond1994,Ishizaki1996,Shinohara1995} and of Si(111)-7\texttimes7 around 700$^{\circ}\mathrm{C}$-780$^{\circ}\mathrm{C}$ \citep{Omi2005}, which both show step bunching under deposition. 
A quasistatic stability analysis using the classical model of step dynamics\footnote{
The ``classical'' model does not include the chemical effect, see e.g., \cite{Pierre-Louis2003}.}
 predicts that the direct Schwoebel effect ($S \ge 1$) and elastic interactions are both stabilizing (see Section~4.2 of Part I and also Table~\ref{tab:dynInfl}). In this context, observations of step bunching under deposition were first explained by invoking an inverse 
 Schwoebel effect ($S<1$) \citep{Ishizaki1996,Tejedor1998}.

However, as noticed by several authors \citep{Pimpinelli2000,Vladimirova2001,Slanina2005}, an inverse ES barrier---favoring attachment of adatoms from the upper terrace---is difficult to justify both experimentally and theoretically. 
In particular, in GaAs(001) and Si(111)-7\texttimes7, the works that intended to measure the ES barrier conclude to the existence of a direct ES barrier ($S>1$) in GaAs(001) while they lead to contradictory results in Si(111). Indeed, for GaAs(001) both atomistic simulations with empirical potentials \citep{Salmi1999} and experimental studies of the formation of islands \citep{Smilauer1995,Krug1997} conclude for a direct ES barrier. On the other hand, for Si(111)-7\texttimes7, observation of the denuded zones around the steps \citep{Voigtlander1995,Rogilo2013}, measure of the decay rates of island and hole under evaporation and deposition  \citep{Ichimiya1996}, and the measure of the distributions over terraces of the nucleated islands under deposition \citep{Chung2002}  lead to inconsistent conclusions, namely a direct \citep{Ichimiya1996}, an inverse  \citep{Chung2002,Rogilo2013}, and the absence of ES barrier \citep{Voigtlander1995}.

These results led theoreticians to consider new mechanisms to account for step bunching under deposition like the coupling between the diffusing precursors and the adatoms for the vapor phase epitaxy of GaAs(001) \citep{Pimpinelli2000} or the fast diffusion of adatoms along steps of Si(111)-7\texttimes7 \citep{Politi2000}. While these mechanisms are plausible, there is not clear evidence that there are indeed the cause behind the observed step bunching. In particular, the step bunching on GaAs(001) is also observed in molecular beam epitaxy deposition experiments \citep{Pond1994}, which occurs without precursors. Hence the coupling with precursors cannot be invoked for interpreting the latter experiment.
We show below in this section how, without resorting to additional mechanisms, a stability analysis of the classical step-flow model which includes the chemical effect and the dynamics effect can explain the existence of step bunching under deposition.

%%%%%%%%%%%%%%%%%%%%%%%%%%%%%%%%%%%%%%%%%%%%%%%%%%%%%%%%%%%%%%%%
\subsection{Experiments on Si(111)-7\texttimes7}
\label{sub:siliconExp}
We start by giving a quick overview of the selection of the material parameters for Si(111)-7\texttimes7 (details can be find in \ref{app:parameters}).  The coefficient of dipole-dipole elastic interactions $\oa$ can be estimated from the work of \citet{Stewart1994}, while noting its strong dependence on the terrace width $L_0$ through $\oa \propto L_0^{-3}$. In the experiments of \citet{Omi2005}, $L_0$ typically varies between $10~\mathrm{nm}$ and $60~\mathrm{nm}$, which corresponds to $\oa$ ranging from $2 \times 10^{-7}$ and $4\times 10^{-4}$. As discussed before, the very nature of the Schwoebel barrier in Si(111)-7\texttimes7 is controversial, hence we assume symmetric attachment/detachment. In the absence of precise information, we consider a low value of adatom coverage $\Theta=0.01$\footnote{Note that $\Theta=0.01$ taken in the low range of possible $\Theta$ is a conservative value, in the sense that, such a low value of $\Theta$ tends to minimize the influence of the chemical and dynamics effects, which we focus on.}. 
The values of $\oF$ and $\ok$ being either difficult to estimate or unknown, we take an intermediate value\footnote{%
The reader might have in mind that changing $F$ essentially changes the relative importance of elasticity (see Section~\ref{sub:inconsistencies}).}%
of $\oF=10^{-2}$  and allow values for $\ok$ that cover the different possible kinetic regimes (more details are given in \ref{app:parameters}).
Finally, in the absence of evidence that the steps of Si(111)-7\texttimes7 are permeable \citep{Chung2002}, we take $\okp=0$.

\begin{figure*}[tb]
\begin{center}
\includegraphics[width=0.9\textwidth]{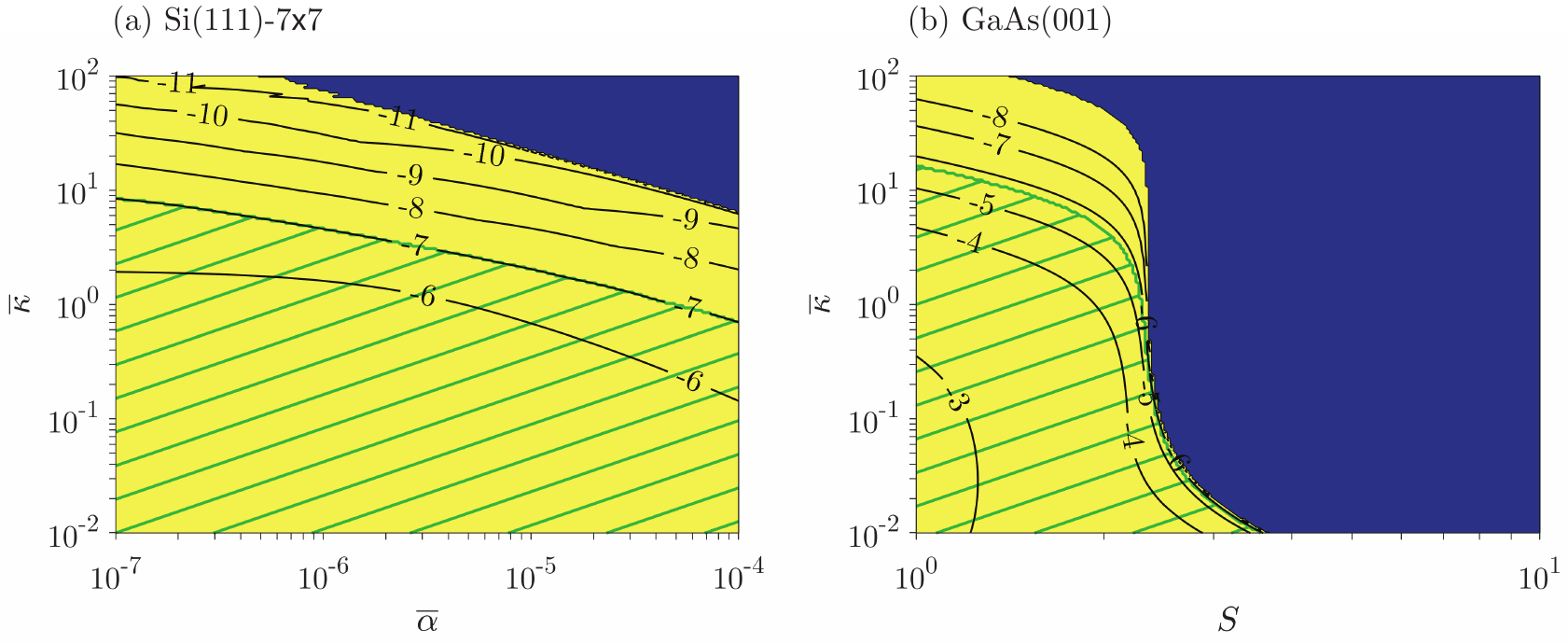}
\end{center}
\caption{(a) Stability diagrams for Si(111)-7\texttimes7 under deposition ($\onu=0$) combining elasticity, chemical and dynamics effects. The parameters are $\Theta=0.01$, $\oF=10^{-2}$, $S=1$, $\ob=0$, $\okp=0$.  (b) Stability diagrams for GaAs(001) under deposition ($\onu=0$) combining elasticity; Schwoebel, chemical, and dynamics effects. The parameters are $\Theta=0.2$, $\oF=10^{-2}$, $\oa=5\times 10^{-6}$, $\ob=0$, $\okp=0$. (see Figure~\ref{fig:elasInfl} for the legends of colors and isolines)}
\label{fig:exp}
\end{figure*}

With the above estimated parameters, and in particular in the absence of Schwoebel barrier, we can see on the stability diagram of Figure~\ref{fig:exp}(a)%
%%%%%%%%%%%%%%%%%%%%%FIGURE 10%%%%%%%%%%%%%%%%%%%%%%%%%%%%%%%
%%%%%%%%%%%%%%%%%%%%%%%%%%%%%%%%%%%%%%%%%%%%%%%%%%%%%%%%%%
%%%%%%%%%%%%%%%%%%%%%%%%%%%%%%%%%%%%%%%%%%%%%%%%%%%%%%%%%%
%%%%%%%%%%%%%%%%%%%%%%%%%%%%%%%%%%%%%%%%%%%%%%%%%%%%%%%%%%
that there exists a large green hatched zone of significant instability. While in this region, bunches develop in a thousand deposited monolayers, the more restrictive condition of one hundred monolayers---for bunches to develop---corresponds to the area below the isoline ``-6''. In this region the speed of bunching is comparable with the observations \cite{Omi2005} in which bunches are observed between 30 and 300 monolayers, thus indicating that the destabilizing mechanisms involved here (chemical and dynamics effect) are sufficiently strong to be a good candidate to explain the instability in these experiments. In sum, the condition for chemical and dynamics effects to induce an instability growing in times comparable to those observed in the experiments of \citet{Omi2005} is that the step kinetics is slow ($\ok<1$). Further experimental work directly assessing the step kinetics would help to identify the actual implication of the proposed mechanism.

%%%%%%%%%%%%%%%%%%%%%%%%%%%%%%%%%%%%%%%%%%%%%%%%%%%%%%%%%%%%%%%%
\subsection{Experiments on GaAs(001)}
\label{sub:GaAsExp}
We move on to the evaluation of the parameters for GaAs(001) at temperatures in the range $600^{\circ} \mathrm{C}-700^{\circ} \mathrm{C}$. Measurements or simulations of the Schwoebel barrier in GaAs(001) suggest a direct Schwoebel effect with estimates (see \ref{app:parameters}) ranging from $S=2$ to $S=10$ at $600^{\circ} \mathrm{C}$. In addition, as detailed in \ref{app:parameters}, the elastic interaction parameter is about $\oa=5 \times 10^{-6}$ and the equilibrium adatom coverage is high, at around $\Theta=0.2$. As discussed in Section~\ref{sub:inconsistencies}, this implies that the chemical and dynamics effects are, compared to materials with a lower adatom coverage, relatively influential. Finally, without any evidence for the existence of permeability in GaAs(001), we take $\okp=0$.

With these parameters, elasticity plays little role and the stability diagram that can be see on Figure~\ref{fig:exp}(b)  results essentially from the combination of the Schwoebel, chemical and dynamics effects. We can see that for $S>2$ the Schwoebel effect prevails, which leads to stable step flow irrespective of the step kinetic regime. By contrast, for $S \leq 2$, a significant step bunching is predicted for slow step kinetics ($\ok < 1$). Hence, for bunching observed on GaAs(001), dynamics and chemical effects furnish plausible explanation. The critical point here is the strength of the Schwoebel barrier whose exact estimation is rendered difficult by the existence of different types of steps and of two simultaneously diffusing species, gallium and arsenic.

%%%%%%%%%%%%%%%%%%%%%%%%%%%%%%%%%%%%%%%%%%%%%%%%%%%%%%%%%%%%%%%%
%%%%%%%%%%%%%%%%%%%%%%%%%%%%%%%%%%%%%%%%%%%%%%%%%%%%%%%%%%%%%%%%
\section{Discussion}
\label{sec:discussion}

\subsection*{Inadequacy of the quasistatic approximation}

The comparison,  \textit{a posteriori}, of the stability results with and without dynamics (see Section~\ref{sub:inconsistencies}) shows the incorrectness of the statement of the quasistatic approximation \citep[as given e.g., in][]{Krug2005,Michely2012} according to which dynamics terms are negligible under the slow deposition/evaporation condition $\oF \Theta \ll 1$ or $\onu \Theta \ll 1$.

Beyond that demonstration by the results, we can see in the stability problem itself that there is no reason for neglecting \textit{a priori} the dynamics terms in the slow deposition/evaporation regimes. Indeed, determining the stability of  the steady-state solution of \eqref{eq:fbvp2} is tantamount to computing, for all wavelengths $k$,  the eigenvalue of \eqref{eq:lpebw} with the largest real part. Yet, in the expressions of the operators $\hat{\mathcal{A}}_k$ and $\hat{\mathcal{B}}_k$ of \eqref{eq:lpebw} given in \ref{app:lsadetails}, we show in red the terms coming from the dynamics terms in \eqref{eq:fbvp2}. Among these terms, we find in addition to terms proportional to $\vz$ (which are of order $\oF \Theta$ under deposition and $\onu \Theta$ under evaporation\footnote{%
Under deposition $\vz=\oF \Theta$, while under evaporation  $\vz = \onu \Theta +o(\onu)$.}),
additional terms of order $0$ (with respect to $\oF \Theta$ and $\onu \Theta$). Hence, even for vanishingly small deposition/evaporation rates, the generalized eigenvalue problem with dynamics terms taken into account differs from its counterpart in the quasistatic regime. This helps us understand why, as shown in Section~\ref{sub:inconsistencies}, stability results with and without dynamics differ significantly.

\subsection*{The different approaches to the effect of dynamics}

In the present work, we have undertaken a thorough analysis of the effect of dynamics by investigating the stability of the complete step-flow problem \eqref{eq:fbvp2}. This follows previous works that considered, at least partially, the effect of dynamics by adopting different approaches. We put these different works in perspective to highlight in particular the existence of two alternative methods to compute the stability of \eqref{eq:fbvp2}.

The work of \citet{Ranguelov2007} addressed the problem of step dynamics in the simplified framework of infinitely fast terrace diffusion ($D \rightarrow \infty$) and slow attachment/detachment kinetics ($\kappa_-$ and $\kappa_+$ small but finite). They also discuss the effect of the dynamics on the problem of step bunching under electromigration \citep{Ranguelov2008} and in another work, show experimental evidence that the dynamics effect may induce step bunching for sufficiently high deposition rates \citep{Ranguelov2017}. The results presented in Figure~\ref{fig:elasInfl} are in qualitative agreement with those of \citet{Ranguelov2007}, noting that the limit case considered by these authors corresponds to $\ok \ll 1$.
However, the problem they consider takes a very different form from the general set of governing equations \eqref{eq:fbvp2}, which makes a quantitative comparison of their analysis with our work difficult.   

In the same decade, other works \citep{Gillet2000,Pierre-Louis2003,Dufay2007} have considered  the possible influence of the dynamics terms on problems similar  to  \eqref{eq:fbvp2}. In works by \citet{Pierre-Louis2003,Dufay2007}, avoiding the mathematical difficulty that the term $\partial_t \rho_n$ in \eqref{eq:fbvp2}$_1$ introduces in  the stability analysis, the authors solely account for the advection part of the dynamics terms%
\footnote{\label{fn:advec}	More precisely, the term $\partial_t \rho_n$ can be decomposed in a convective component  and a  transient component  by doing the change of variable $\tilde{x}:=x-(n+\vz t)$. Letting $\rt_n(\tilde{x},t):=\rho_n(\tilde{x}+n+\vz t,t)$ the time derivative of $\rho_n$ is rewritten $\partial_t \rho_n= \partial_t \rt_n-\vz \partial_x \rt_n$. In the work mentioned here,  only the advective term $-\vz \partial_x \rt_n$ is conserved while the time derivative $\partial_t \rt_n$ is neglected in \eqref{eq:fbvp2}$_1$. Similarly, for the boundary conditions \eqref{eq:fbvp2}$_{2,3}$, the terms $-\rho_n^- \dot{x}_{n+1}$ and $\rho_n^+ \dot{x}_{n}$ are replaced by $-\rho_n^- \vz$ and $\rho_n^+ \vz$ whereby neglecting there the perturbations in the step positions about the principal velocity $\vz$.}. While it seems to us that this approach provides the good stability results for $\oF,\onu \rightarrow 0$, we find several cases, still in the regime of slow deposition rate, where the stability results derived with that simplification differ significantly from the general ones. Indeed, in the absence of clear asymptotics showing that the transient terms can be neglected, our results advocate for keeping all dynamics terms without doing unjustified simplifications.

A treatment of the effect of dynamics on the step stability problem---that include the transient term in \eqref{eq:fbvp2}$_1$--- is developed in \citet{Ghez1990,Ghez1993,Keller1993}. In this work, the authors write the perturbation equations on the domain of the steady-state solutions and correct the inadequacy of the domain of definition through Taylor expansions of the boundary conditions about the steady-state position of the interface. Using the vocabulary used for stability problems in fluid-structure interaction---where the same issue of both function and domain perturbation arises---their approach is referred to as the \textit{transpiration method} by contrast with ours makes use of the \emph{arbitrary Lagrangian-Eulerian formulation}  \citep[see e.g.,][]{Fanion2000}. 
However, in the formulation of the step governing equations by \citet{Ghez1990,Ghez1993,Keller1993}, the dynamics terms are missing from the boundary conditions  \eqref{eq:fbvp2}$_{2,3}$ as those were only included in the step-flow problem at later times \citep[see e.g.,][]{Pierre-Louis2003,Ranguelov2007,Dufay2007}.

More recently, making use again of the transpiration method, \citet{Gillet2000} addresses in his thesis the stability of steps with proper account for all the dynamics terms. However, the resulting effect of dynamics on stability is not discussed much in this work as \citet{Gillet2000} simply notes that the dynamics effect may destabilize the steps under deposition. To fill that gap we have provided, in Section~\ref{sub:lsa2results}, a detailed discussion of the effect of dynamics and of its importance relative to the other stabilizing/destabilizing mechanisms. In particular, we point out there the fundamental role of $\ok$ for understanding the effect of dynamics on stability. 

To conclude this discussion, we mention the work of \citet{Sekerka1967} which is concerned by the stability of a planar solidification front. Although that problem is different from ours, it shares similarities in the formalism, being also a Stefan-like problem. In this work, the author addresses the influence of the equivalent to our ``dynamics terms'' on the stability, however it seems that, like in the work of \citet{Ghez1990} on the step-flow problem, the terms related to the advective current are missing from the boundary conditions. This opens the road for revisiting the effect of dynamics on stability in other problems.

\section{Conclusion}
\label{sec:conclusion}

We have investigated the onset of the bunching instability on vicinal surfaces without recourse to the quasistatic approximation, thereby unraveling the effect of dynamics. We found that both under deposition and evaporation, dynamics can be stabilizing or destabilizing depending on the value of  $\ok$ (expressing the ratio of step attachment/detachment kinetics to terrace diffusion kinetics). In examining how the strength of the dynamics effect varies with the material and operational parameters, we found that the effect of dynamics on stability scales linearly with the deposition/evaporation rates (like for the Schwoebel and chemical effects) and quadratically with the adatom coverage (alike the chemical effect). 

When combined with the other stabilizing/destabilizing mechanisms, we find that dynamics significantly modifies the diagrams of stability and that its influence on stability does not necessarily decreases for small deposition/evaporation rate. Whereas from  a theoretical perspective this shows the inadequacy of the quasistatic approximation as usually invoked in the literature, more practically this calls for reinterpreting the cause of step bunching observed in some experiments. 

With this regards, we combine several experimental works to estimate the physical parameters of the problem and find that dynamics provides a plausible explanation for some cases of step bunching observed in GaAs(001) and Si(111)-7\texttimes7. These explanations appear as an alternative to the additional physical mechanisms invoked previously to account for step bunching. To determine with better certainty the mechanisms indeed involved in particular occurences of bunching, one need to consider the subsequent evolution of the bunches. Such an analysis has been sketched in \citet{Guin2020a} and will be the object of upcoming publications \citep{Benoit2020}.

\section*{Acknowledgment}

This work is supported by the ``IDI 2015'' project funded by the IDEX Paris-Saclay under grant ANR-11-IDEX- 0003-02. 
The authors thank L. Benoit-Maréchal for fruitful discussions.

\newpage
\appendix
%%%%%%%%%%%%%%%%%%%%%%%%%%%%%%%%%%%%%%%%%%%%%%%%%%%%%%%%%%%%%%%%
%%%%%%%%%%%%%%%%%%%%%%%%%%%%%%%%%%%%%%%%%%%%%%%%%%%%%%%%%%%%%%%%
\section{Expressions of the linear stability operators}
\label{app:lsadetails}

\subsection*{Expressions for the operators $\mathcal{A}$ and $\mathcal{B}$}
The linear operator $\mathcal{A}$ introduced in \eqref{eq:lpe} 
is defined by
\begin{multline} \label{eq:aop}
\mathcal{A}\big(\mathbf{q}_{n-1},\mathbf{q}_n,\mathbf{q}_{n+1},\mathbf{q}_{n+2} \big) =  \begin{pmatrix}
A_1^1(u) \, (\dx_n  -\dx_{n+1}) \\
A_2^1  \, \dx_n + A_2^1 {}' \, \dx_{n+1} + A_2^1 {}'' \, \dx_{n+2} \\
A_3^1 \, \dx_{n-1} + A_3^1 {}' \, \dx_{n} + A_3^1 {}'' \, \dx_{n+1} \\
A_4^1 \, ( \dx_{n-1} -2 \dx_{n} + \dx_{n+1} )
\end{pmatrix} + \begin{pmatrix}
A_1^2\, \drt_n(u,t)  \\
A_2^2\, \drt_{n+1}^+ + A_2^2 {}' \, \drt_n^- \\
A_3^2\, \drt_n^+ + A_3^2 {}' \, \drt_{n-1}^- \\
A_4^2\, \drt_n^+ + A_4^2 {}' \, \drt_{n-1}^- 
\end{pmatrix} \\
+ \begin{pmatrix}
\vz \partial_u \drt_n(u,t) \\
 (\partial_u \drt_n)^- \\
- (\partial_u \drt_n)^+ \\
0
\end{pmatrix} + \begin{pmatrix}
\partial_{uu} \drt_n(u,t) \\
0 \\
0 \\
0
\end{pmatrix},
\end{multline}
 where
 \begin{empheq}[left=\empheqlbrace]{align}
\begin{aligned}
A_1^1(u) = &\,  2\nu \rtz(u) - 2F -\vz  \rtzp(u), \\
A_2^1  = & \,\ok (1+3 \oa - \ob ) + (\okp + \ok \chi \Theta ) \rtz(0) - \big ( \ok (1+\chi \Theta)+\okp+\vz \big) \rz(1), \\
A_2^1 {}' = &\, - \ok (1+6 \oa - 2\ob ) - (\okp + \ok \chi \Theta ) \rtz(0) +\big ( \ok (1+\chi \Theta)+\okp+\vz \big) \rz(1), \\
A_2^1 {}'' =&\, \ok (3 \oa - \ob ), \quad A_3^1 = \ok S (3 \oa - \ob ), \\
A_3^1 {}' = &\, \ok S (1-6 \oa +2 \ob ) + (\okp - \ok S \chi \Theta ) \rtz(1)  + \big ( \ok S+ (\chi \Theta-1)-\okp+\vz \big) \rz(0), \\
A_3^1 {}'' = & \,- \ok S (1-3 \oa + \ob ) - (\okp - \ok S \chi \Theta ) \rtz(1)  + \big ( \ok S+ (1-\chi \Theta)+\okp-\vz \big) \rz(0), \\
A_4^1 = & \, \Theta \ok(1+S)(\ob - 3 \oa), \\
A_1^2 = & \,- \onu, \quad A_2^2 = - \ok \chi \Theta - \okp, \quad A_2^2 {}' = \ok (1+\chi \Theta)+\okp+\vz, \\
A_3^2 = &\, \ok S(1-\chi \Theta)+\okp-\vz, \quad A_3^2 {}' = \ok S \chi \Theta - \okp, \\
A_4^2 = &\, \Theta \ok \big(\chi \Theta (1+S)-S\big), \quad A_4^2 {}' = - \Theta \ok \big(\chi \Theta (1+S)+1\big). \\
\end{aligned}
\end{empheq}
The linear operator $\mathcal{B}$ also introduced in \eqref{eq:lpe} 
is defined by
\begin{equation} \label{eq:bop}
\mathcal{B} \big( \partial_t \mathbf{q}_n,\partial_t \mathbf{q}_{n+1} \big) := \begin{pmatrix}
(u-1) \rtzp(u) \ddx_n - u \rtzp(u) \ddx_{n+1} \\
\rtz(1) \ddx_{n+1} \\
\rtz(0) \ddx_{n} \\
\ddx_{n}
\end{pmatrix} + \begin{pmatrix}
\partial_t \drt_n(u,t) \\
0 \\
0 \\
0
\end{pmatrix}.
 \end{equation}

\subsection*{Expression of the operators $\hat{\mathcal{A}}_k$ and $\hat{\mathcal{B}}_k$}
These operators are obtained
as explained in Section~\ref{sub:lsa2}, 
by inserting \eqref{eq:bw} in \eqref{eq:lpe}. The operator $\hat{\mathcal{A}}_k$ reads\footnote{In red are the terms coming from dynamics, see the discussion in Section~\ref{sec:discussion}.}
\begin{equation}
 \begin{aligned}
\label{eqn:DecompositionAk} \hat{\mathcal{A}}_k \hat{\mathbf{q}} & = = \begin{pmatrix}
\ha_1^1\,(u) \dxh \\
\ha_2^1\, \dxh \\
\ha_3^1\, \dxh \\
\ha_4^1\, \dxh 
\end{pmatrix} + \begin{pmatrix}
\ha_1^2\, \drh(u)  \\
\ha_2^2\, \drh(0) + \ha_2^2 {}'\, \drh(1) \\
\ha_3^2\, \drh(0) + \ha_3^2 {}'\, \drh(1) \\
\ha_4^2\, \drh(0) +  \ha_4^2 {}'\, \drh(1) 
\end{pmatrix} + \begin{pmatrix}
\ha_1^3\, \drh'(u) \\
\ha_2^3\, \drh'(1) \\
\ha_3^3\, \drh'(0) \\
0
\end{pmatrix} + \begin{pmatrix}
\ha_1^4\, \drh''(u) \\
0 \\
0 \\
0
\end{pmatrix},
\end{aligned}
\end{equation}
with
\begin{empheq}[left=\empheqlbrace]{align}
\begin{aligned}
\ha_1^1(u)=& \left(-1+\mathrm{e}^{i k}\right) \left(2 \oF-2 \onu \rhz(u)+\cb{\vz \rhz'(u)}\right), \\
\ha_2^1=&\left(-1+\mathrm{e}^{i k}\right) \bigg(\ok \left((\mathrm{e}^{i k}-1) (3 \oa-\ob)+(\rhz(1)-\rhz(0)) \chi \Theta+\rhz(1)-1\right)\\
& \qquad \qquad \qquad \qquad \qquad \qquad \qquad  + \rhz(1) (\okp+\cb{\vz})-\okp \rhz(0)\bigg), \\
\ha_3^1=&\left(-1+\mathrm{e}^{i k}\right) \bigg(\ok S \left(  (1-\mathrm{e}^{-i k} )(3 \oa-\ob)-\rhz(0) (\chi \Theta-1)+\rhz(1) \chi \Theta-1\right) \\
& \qquad \qquad \qquad \qquad \qquad \qquad  \qquad  + \rhz(0)( \okp-\cb{\vz})  -\okp \rhz(1)\bigg),\\
\ha_4^1=&-2 \ok (S+1) \chi \Theta (3 \oa-\ob) (\cos (k)-1), \\
\ha_1^2=&- \onu, \quad \ha_2^2=-\mathrm{e}^{i k} (\ok \chi \Theta+\okp), \quad \ha_2^2 {}'=\ok + \okp + \ok \chi \Theta + \cb{\vz}, \\
\ha_3^2=&-\ok S \chi \Theta+\ok S+\okp-\cb{\vz}, \quad \ha_3^2 {}'=\mathrm{e}^{-i k} (\ok S \chi \Theta-\okp), \\
\ha_4^2=&\ok  \Theta \big((S+1) \chi \Theta-S\big), \quad \ha_4^2 {}'=-\mathrm{e}^{-i k} \ok \Theta \big((S+1) \chi \Theta+1\big), \\
\ha_1^3=&\cb{\vz}, \quad \ha_2^3=1, \quad \ha_3^3=-1, \quad \ha_1^4= 1. \\
\end{aligned}
\end{empheq}
The corresponding expressions for the operator $\hat{\mathcal{B}}_k$ is
\begin{equation}
\begin{aligned}
\label{eqn:DecompositionBk} \hat{\mathcal{B}}_k \hat{\mathbf{q}}  =  \begin{pmatrix}
\hb_1^1\,(u) \dxh \\
\hb_2^1\, \dxh \\
\hb_3^1\, \dxh \\
\hb_4^1\, \dxh 
\end{pmatrix} + \begin{pmatrix}
\hb_1^2\, \drh(u) \\
0 \\
0 \\
0 
\end{pmatrix},
\end{aligned}
\end{equation}
with
\begin{empheq}[left=\empheqlbrace]{align}
\begin{aligned}
\hb_1^1(u)=&\cb{-\left(1+\left(-1+\mathrm{e}^{i k}\right) u\right) \rhz'(u)}, \quad \hb_2^1=\cb{-\me^{i k} \rhz(1)}, \\
\hb_3^1=&\cb{\rhz(0)}, \quad \hb_4^1=-1, \quad \hb_1^2=\cb{1}.
\end{aligned}
\end{empheq}
%
%%%%%%%%%%%%%%%%%%%%%%%%%%%%%%%%%%%%%%%%%%%%%%%%%%%%%%%%%%%%%%%%

%%%%%%%%%%%%%%%%%%%%%%%%%%%%%%%%%%%%%%%%%%%%%%%%%%%%%%%%%%%%%%%%
\section{Numerical method}
\label{app:numerical}

The eigenvalue problem \eqref{eq:lpebw} involves two operators $\hat{\mathcal{A}}_k$ and $\hat{\mathcal{B}}_k$ acting on the scalar variable $\delta \hat{x}$ and on the function $\delta \hat{\rho} \, : \, [0,1] \rightarrow \mathbb{C}$ and its derivatives.  This problem is solved numerically, using the Chebyshev collocation method \citep{peyret2002}, a pseudo-spectral method adapted to nonperiodic problems.

\begin{figure*}[tb]
\centering
\includegraphics[width=0.7\textwidth]{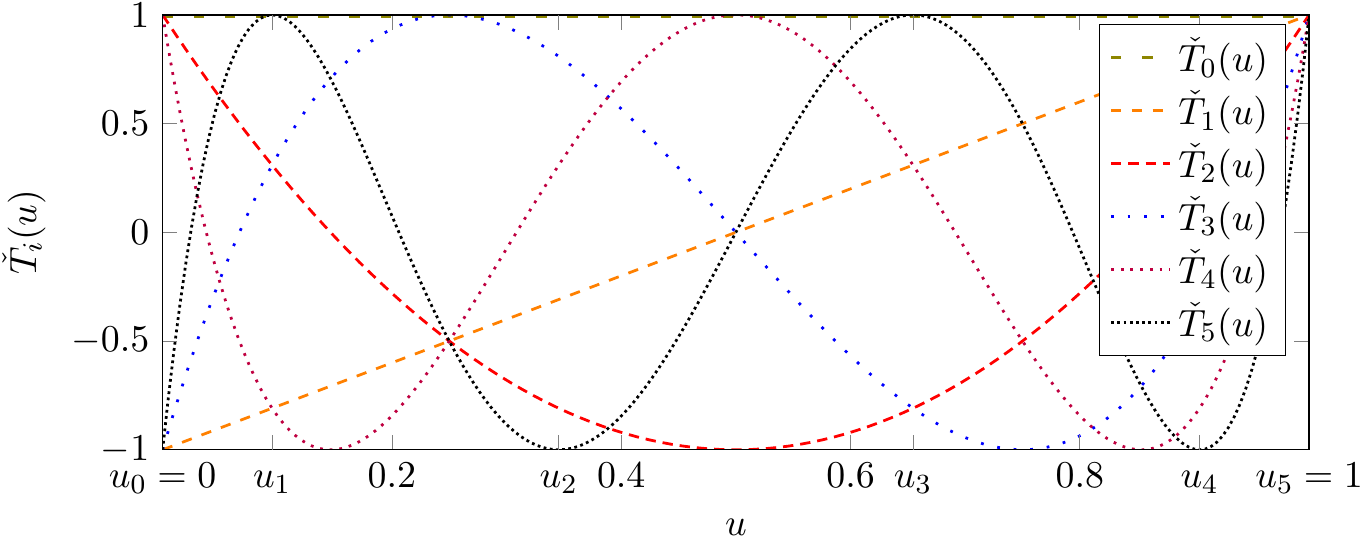}
\caption{Representation of the first six Chebyshev polynomials $\check{T}_n$ along with the six Gauss-Lobatto points for $N=5$.}
\label{fig:PolynomesChebyshev} 
\end{figure*}

Consider a complex-valued function $f: u \in [0,1] \mapsto f(u) \in \mathbb{C}$  and let its Chebyshev series approximation $f_N$ truncated at order $N$,
\begin{equation}
 f_N(u) := \sum_{n=0}^N \check{f}_n \check{T}_n(u),
\end{equation}
where $\check{f}_n$ are the Chebyshev coefficients and  $\check{T}_n$ denote the Chebyshev polynomials of the first kind\footnote{Note that the Chebyshev polynomials of the first kind $\check{T}_n$ defined on $[0,1]$ used here are obtained by rescaling their traditional counterparts $T_n$ defined on $[-1,1]$ via
$\check{T}_n (u) := T_n \left( 2u - 1 \right), \ \forall u \in [0,1].$}, depicted on Figure~\ref{fig:PolynomesChebyshev}%
%
%%%%%%%%%%%%%%%%%%%%%FIGURE 01%%%%%%%%%%%%%%%%%%%%%%%%%%%%%%%
%%%%%%%%%%%%%%%%%%%%%%%%%%%%%%%%%%%%%%%%%%%%%%%%%%%%%%%%%%
%%%%%%%%%%%%%%%%%%%%%%%%%%%%%%%%%%%%%%%%%%%%%%%%%%%%%%%%%%
%%%%%%%%%%%%%%%%%%%%%%%%%%%%%%%%%%%%%%%%%%%%%%%%%%%%%%%%%%
 for $n=0\ldots5$,
 and defined on $[0,1]$ by
the recurrence relationship
\begin{equation}
\check{T}_0(u)=1, \quad \check{T}_1(u)=2u-1,\quad \check{T}_{n}=(4u-2)\check{T}_{n-1}-\check{T}_{n-2}=0 \quad \mbox{for} \; n \geq 2.
\end{equation}
To compute the Chebyshev coefficients $\check{f}_n$, we use the collocation method on the Gauss-Lobato points $u_0, \ldots, u_N$ defined as
\begin{equation}
u_n := \frac{1}{2}  \left[ \cos \left( \frac{\pi (N-n)}{N} \right)+1 \right], \quad 0 \le n \le N,
\end{equation}
and shown on Figure~\ref{fig:PolynomesChebyshev} for $N=5$.

With the Chebyshev approximation, the first $f'$ and second $f''$ derivatives of $f$ are approximated by
\begin{equation}
f'_N(u) = \sum_{n=0}^N \check{f}_n \check{T}'_n(u), \quad  f''_N(u) = \sum_{n=0}^N \check{f}_n \check{T}''_n(u).
\end{equation}
The advantage of using the collocation method to compute the Chebyshev coefficients $\check{f}_n$ stems from the fact that the values of $f_N'$ and $f_N''$ at the collocation points $u_n$ can be 
obtained using a fixed differentiation matrix $\mathbf{\check{D}}$ operating on the values of the function on the Gauss-Lobatto points 
$f_N (u_j)$\footnote{Note that the coefficients of the $\mathbf{\check{D}}$ are obtained by adapting the differentiation matrix $\mathbf{D}$ found in \citet{peyret2002} (which corresponds to Chebyshev polynomials defined on $[-1,1]$) to the Chebyshev method reformulated on $[0,1]$.}:
\begin{equation}
f_N' (u_n) = \sum_{j=0}^N \check{D}_{n j} f_N (u_j).
\end{equation}
Note that, with the Chebyshev method, the derivative at one point does not only depend on the neighboring points but on all the points of the domain which makes $\mathbf{\check{D}}$ a full matrix. Similarly, the second-order derivative is approximated using $(\check{\mathbf{D}})^2$.

To derive the discrete form of \eqref{eq:lpebw}, the operators $\hat{\mathcal{A}}_k$ and $\hat{\mathcal{B}}_k$  are decomposed as a sum of operators acting separately on $\dxh$, $\drh$ and its derivatives:
\begin{equation} \label{eq:decompo}
\hat{\mathcal{A}}_k (\dxh,\drh) = \hat{\mathcal{A}}_k^1 \dxh + \hat{\mathcal{A}}_k^2 \drh + \hat{\mathcal{A}}_k^3 \drh' + \hat{\mathcal{A}}_k^4 \drh'' ,\quad
 \hat{\mathcal{B}}_k (\dxh,\drh) = \hat{\mathcal{B}}_k^1 \dxh + \hat{\mathcal{B}}_k^2 \drh,
\end{equation}
where the full expressions of the $\hat{\mathcal{A}}_k^p$, $p=1\ldots4$ and $\hat{\mathcal{B}}_k^m$, $m=1,2$ are given in \ref{app:lsadetails}.

Using the decomposition \eqref{eq:decompo}, the operators $\hat{\mathcal{A}}_k$  and $\hat{\mathcal{B}}_k$ are approximated using the  $(N+2) \times (N+2)$ matrices $\check{\mathbf{A}}^k$ and $\check{\mathbf{B}}^k$ written as
\begin{equation}
\check{\mathbf{A}}_k \check{\mathbf{q}} =  \check{\mathbf{A}}^1_k \dxh  + \check{\mathbf{A}}^2_k \drb + \check{\mathbf{A}}^3_k  \check{\mathbf{D}}  \drb + \check{\mathbf{A}}^4_k   (\check{\mathbf{D}})^2 \drb, \quad
\check{\mathbf{B}}_k \check{\mathbf{q}} =  \check{\mathbf{B}}^1_k \dxh + \check{\mathbf{B}}^2_k \drb,
\end{equation}
where $\drb= \big(\drh (u_0), \ldots,\drh (u_N) \big)$, $\check{\mathbf{q}} =  (\dxh, \drb)$ and $\check{\mathbf{A}}_k^p$, $p=1\ldots4$ and $\check{\mathbf{B}}_k^m$, $m=1,2$  are the discretization of the corresponding operators $\hat{\mathcal{A}}_k^p$ and $\hat{\mathcal{B}}_k^m$ on the Gauss-Lobatto mesh $\{ u_i, 0 \leq i \leq N \}$. 

This allows to discretize \eqref{eq:lpebw} in a $(N+2) \times (N+2)$ generalized eigenvalue problem: For a given $k \in [0,\pi]$, find $(\lambda, \check{\mathbf{q}}) \in (\mathbb{C} \times \mathbb{C}^{N+2})$ with $\check{\mathbf{q}} \neq \mathbf{0}$ such that 
\begin{equation} \label{eq:lpediscreet}
\check{\mathbf{A}}_k \check{\mathbf{q}} = \lambda  \check{\mathbf{B}}_k \check{\mathbf{q}}.
\end{equation}
After solving \eqref{eq:lpediscreet} numerically, we consider the leading eigenvalue $\lambda$  (the eigenvalue with largest real part), which corresponds to the most critical growth rate.

As  $\check{\mathbf{B}}_k$ is not invertible, \eqref{eq:lpediscreet} has less than $N+2$ eigenvalues $\lambda$. Indeed as can be deduced from \eqref{eqn:DecompositionBk} in \ref{app:lsadetails}, the last three rows of $\check{\mathbf{B}}_k$ being linearly dependent, its kernel is of dimension 2. As a result, and noting that $\check{\mathbf{A}}^k$ is invertible, \eqref{eq:lpediscreet} has only $N$ eigenvalues.

\subsection*{Convergence}

We evaluate the convergence of the numerical method by considering the leading eigenvalue, and determine a satisfactory number of discretization points $N+1$. 

For a given set of parameters ($\oF=10^{-2}$, $\onu=0$, $S=1$, $\ok=10$, $\okp=0$, $\Theta=0.01$, $\oa=0$, $\ob=0$ and $k=\pi/2$), the leading eigenvalue with $N=50$ is $-1.1271 \cdot 10^{-7} + 1.0000 \cdot 10^{-4} i$. Because of the spectral convergence properties of Chebyshev methods \citep{peyret2002}, full convergence with five significant digits of the leading eigenvalue is achieved with $N$ as low as $5$. This fast convergence has been verified for various set of parameters and we select the value $N=10$ for the present analysis.

\begin{figure*}[tb]
\centering
\includegraphics[width=0.7\textwidth]{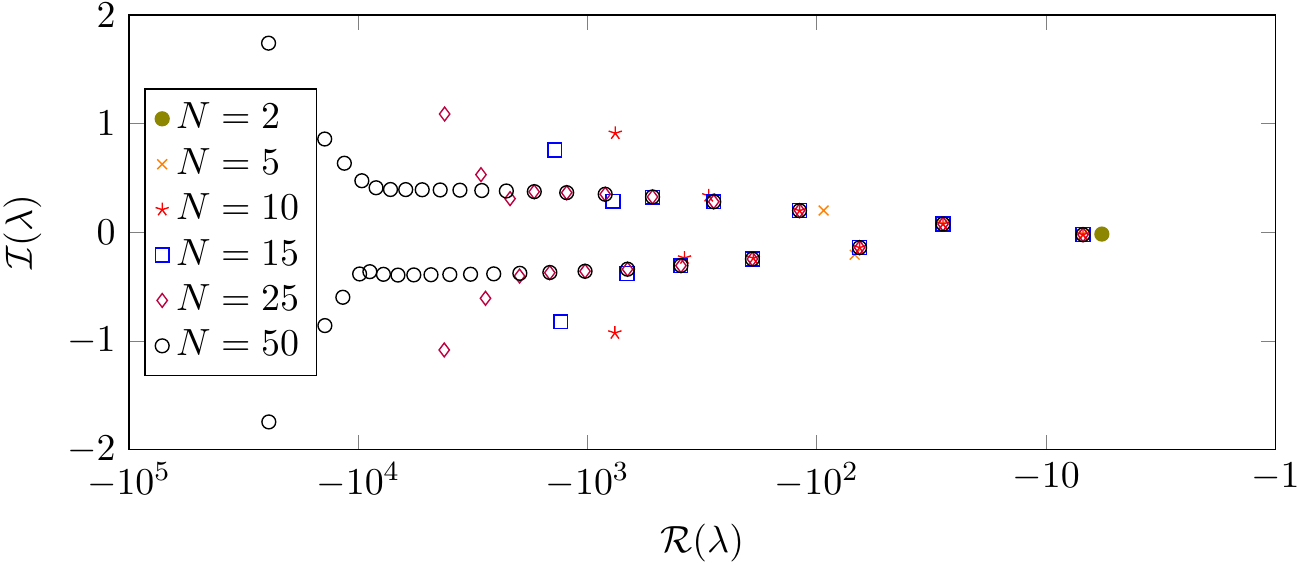}
\caption{Evolution of the eigenvalue spectrum of \eqref{eq:lpediscreet}	with the number $N$ of discretization points for the set of parameters $\oF=10^{-2}$, $\onu=0$, $S=1$, $\ok=10$, $\okp=0$, $\Theta=0.01$, $\oa=0$, $\ob=0$ and $k=\pi/2$.  The logarithmic scale used on the abscissa is not wide enough to include the leading eigenvalue $-1.1271 \cdot 10^{-7} + 1.0000 \cdot 10^{-4} i$.}
\label{fig:ConvergenceSpectreN} 
\end{figure*}

A superposition of eigenvalue spectra for different values of $N$ is shown in Figure~\ref{fig:ConvergenceSpectreN}.%
%%%%%%%%%%%%%%%%%%%%%FIGURE 02%%%%%%%%%%%%%%%%%%%%%%%%%%%%%%%
%%%%%%%%%%%%%%%%%%%%%%%%%%%%%%%%%%%%%%%%%%%%%%%%%%%%%%%%%%
%%%%%%%%%%%%%%%%%%%%%%%%%%%%%%%%%%%%%%%%%%%%%%%%%%%%%%%%%%
%%%%%%%%%%%%%%%%%%%%%%%%%%%%%%%%%%%%%%%%%%%%%%%%%%%%%%%%%%
 On top of the leading eigenvalue previously discussed, we observe the progressive formation of two branches of constant imaginary part in the stable spectral plane. Note that, although these branches are not of particular interest for the linear stability analysis, as expected from a consistent numerical method they get more and more resolved as $N$ increases.

%************************************************
\section{Material parameters of crystal growth}
\label{app:parameters}
%************************************************

Based on the material properties of GaAs and Si found in the experimental literature, we select here the values of the dimensionless parameters of the step-flow problem \eqref{eq:fbvp2}. When no measured values of these parameters are available, we provide physically plausible ranges.

%%%%%%%%%%%%%%%%%%%%%%%%%%%%%%%%%%%%%%%%%%%%%%%%%%%%%%%%%%%%%%%%
\subsection*{Upper bounds on $\oF$ and $\onu$}

Before referring to experimental works, we first note that the satisfaction of the ``near equilibrium'' assumption ($\vert \rho_n(x,t) - \rho_{eq}^* \vert \ll \rho_{eq}^*$ with dimensional quantities) underlying the derivation of \eqref{eq:fbvp2} (see Section 2.1 of the companion paper) implies upper bounds on $\oF$ and $\onu$. Physically, given the finite diffusion speed of adatoms, a high deposition (evaporation) rate may lead to adatom density on the terraces violating the ``near equilibrium'' assumption. 

For estimating these bounds, consider first the deposition case ($\onu =0$) with equidistant steps, under the assumption of infinite a/d velocities ($\ok \rightarrow \infty$). This simplifying assumption---which tends to underestimate the adatom density on terraces---implies that  $\rz^\pm=1$ and $\rhz(x)=1-1/2 \oF x (x-1)$ with a maximum adatom densities of $\rz_{max}=1+\oF/8$ at $x=1/2$. Hence, limiting adatom density variations to, for example, one fourth of the equilibrium value, implies the restriction $\oF<2$. A similar reasoning on the evaporation case ($\oF=0$) implies that $\onu <2.5$.

 In sum, in order of magnitude, the compliance with the near equilibrium assumption requires $\oF$ and $\onu$ not to be more than unity. This condition is related---and actually more stringent---than the condition of sufficiently slow deposition/evaporation rate ($\oF \Theta< 1$ and $\onu \Theta< 1$, see \citet{Krug2005,Michely2012}), which ensures that the crystal grows in the step-flow regime and not through island nucleation and growth.

%%%%%%%%%%%%%%%%%%%%%%%%%%%%%%%%%%%%%%%%%%%%%%%%%%%%%%%%%%%%%%%%

%%%%%%%%%%%%%%%%%%%%%%%%%%%%%%%%%%%%%%%%%%%%%%%%%%%%%%%%%%%%%%%%
\subsection*{Equilibrium adatom coverage $\Theta$}

The equilibrium adatom coverage for GaAs(001) and Si(111) has been measured by different groups, with the same technique consisting in rapid quenching of an equilibrium vicinal surface and observations of islands resulting from the crystallization of adatoms. 

For GaAs(001), \citet{Johnson1996} \citep[see also ][]{Johnson1997,Tersoff1997} measured the equilibrium adatom coverage under typical Molecular Beam Epitaxy (MBE) conditions and found values between $\Theta=0.05$ and $\Theta=0.2$ for temperatures between $570\dc$ and $600\dc$.

For Si(111) we haven't find any data on the adatom coverage in the low temperature regime ($650 \dc$ to $850 \dc$) where silicon exhibits a $7\! \times \! 7$ surface reconstruction and hence assume a low value of $\Theta=0.01$.

%%%%%%%%%%%%%%%%%%%%%%%%%%%%%%%%%%%%%%%%%%%%%%%%%%%%%%%%%%%%%%%%
\subsection*{Kinetic a/d coefficient $\ok$ and kinetic permeability coefficient $\okp$}

The ratio of attachment/detachment (a/d) kinetics to surface diffusion kinetics, reflects an a/d limited (ADL) regime when $\ok \ll1$ and a diffusion limited (DL) regime when $\ok \gg 1$. In general, given the indeterminacy of the kinetic regime, we cover both regimes by considering values of $\ok$ ranging from $10^{-2}$ to $10^2$. 

More specifically, for Si(111)-7\texttimes7 it seems that the kinetic regime changes from ADL at low temperatures \citep[i.e., $400\dc-600\dc$ see][]{Ichimiya1996} to DL at high temperature \citep[i.e., $\sim 860\dc$ see][]{Hibino2001}, however we have no direct information on the step kinetics in the temperature range $700\dc-800\dc$ corresponding to the bunching experiments \citep{Omi2005} which we discuss in Section~\ref{sub:siliconExp}.

\subsection*{Ehrlich-Schwoebel effect $S$}

The Ehrlich-Schwoebel (ES) effect (quantifying the asymmetry of the a/d coefficients at the steps and defined by $S:= \kappa_{+}/\kappa_{-}$) is also rarely clearly determined experimentally. In absence of accurate information on the Schwoebel barrier, we explore here both direct and inverse ES effect with a maximum ratio of the upper and lower attachment coefficients of an order of magnitude.

\paragraph{ES barrier in Si(111)-7\texttimes7}

Considering, Si(111)-7\texttimes7, measurements of the Schwoebel effect have been performed by different techniques---comparison in the growth and decay rates of islands and holes \citep{Ichimiya1996}, denuded zones around steps \citep{Voigtlander1995,Rogilo2013}, island nucleation distributions \citep{Chung2002}---and lead to contradictory conclusions, \ie, a direct, negligible or inverse ES effect. Consequently, we use the general range of 
$ 0.1  \le S \le 10$.

\paragraph{ES barrier in GaAs(001)}

For the surface of GaAs(001), different studies yield values for the Ehrlich-Schwoebel barrier in reasonable agreement. By analyzing the characteristics of mounds in the epitaxial growth of GaAs \citet{Smilauer1995} and \citet{Krug1997}  derived values of the direct Schwoebel energy barrier of $\Delta E_-^{Sm}=0.175~\mathrm{eV}$ and $\Delta E_-^{Kr}=0.06~\mathrm{eV}$, respectively. \citet{Krug1997} noted that their method  underestimates---while the one of \citet{Smilauer1995} overestimates---the value of the barrier so that these values should be considered as bounds.   On the other hand, by performing atomistic computations that distinguish between two types of steps on the GaAs(001) surface (parallel and perpendicular to the arsenic dimers), \citet{Salmi1999} obtain a value of the barrier of $\Delta E_-^{Sa}=0.25~\mathrm{eV}$ for one type of step and conclude that there is no barrier for the second type of step.

We use here the intermediate value $\Delta E_-=0.1~\mathrm{eV}$, which can be translated in terms of the parameter $S$ using the relation 
$S=\exp(\Delta E_- /k_B T)$ between the a/d kinetic coefficients and the energetic barriers associated to the hopping of adatoms between upper
and lower terraces \citep{Jeong1999}.
For the temperature of  600$^\circ \mathrm{C}$ at which deposition is considered in  Section~\ref{sec:exp}, this yields $S=4$. Note for estimation of the typical uncertainty, that the parameters $S$ associated with the energetic barriers of \citeauthor{Krug1997} and \citeauthor{Smilauer1995} are $S^{Kr}=2$ and $S^{Sm}=10$, respectively.

%%%%%%%%%%%%%%%%%%%%%%%%%%%%%%%%%%%%%%%%%%%%%%%%%%%%%%%%%%%%%%%%
\subsection*{Dipole-dipole  elastic interaction coefficient $\oa$}

Recall that the dimensionless  elastic interaction coefficient is $\oa=\alpha a^2/({k_B T L_0^3})$, where the dipole strength $\alpha$ is given by $\alpha= {4(1-\nu^2)(d_x^2+d_z^2)}/({\pi E})$, depending on the strength of the dipole $(d_x\,, \, d_z)$, representing the elastic field created by the steps in homoepitaxy.

\paragraph{Estimation of $\oa$ in Si(111)-7\texttimes7}

For homoepitaxy, $\oa$ can be accurately estimated in Si(111)-7\texttimes7 from the work of \citet{Stewart1994}. Combining experimental measurements of the displacement field of a step with simulations, they determined, for a Si(111)-7\texttimes7 step, the normal and tangential dipole moments: $d_z=0.6~ \textrm{eV/\AA}$ and $d_x=1.5~ \textrm{eV/{\AA}}$. With the effective isotropic elastic properties of silicon $E=166~\mathrm{GPa}$ and $\nu=0.2$ \citep{Stewart1994}, we derive the elastic interaction coefficient $\alpha=3~ \textrm{eV.\AA}$ and its dimensionless counterpart at $1000~\mathrm{K}$ (for the reference terrace width $L_0=20~\mathrm{nm}$) as $\oa=5 \times 10^{-5}$. Note that since $\oa \propto L_0^{-3}$, if the initial terrace width is multiplied by two, $\oa$ is decreased by an order of magnitude. Hence, $\oa$ may vary over several decades and elasticity may have small or large effect on the stability depending on the initial miscut angle. 

\paragraph{Estimation of $\oa$ in GaAs(001)}

The interactions between steps of GaAs(001) has recently been studied with \textit{ab initio} computations (specifically Density Functional Theory, DFT) by \citet{Magri2014,Magri2016}. Through fitting of the atomic displacement field, they derived elastic dipole moments of the order of $0.1~\textrm{eV/\AA}$ and a resulting elastic interaction coefficient of about $\alpha=0.2~\textrm{eV.\AA}$ (the exact value depending on the structure of each step). For an initial terrace width  $L_0=16~\mathrm{nm}$ (corresponding to a miscut angle of $1^\circ$ with step height $0.28~\mathrm{nm}$), the dimensionless value of $\alpha$ is at $1000~\mathrm{K}$: $\oa=5\times 10^{-6}$. 

%%%%%%%%%%%%%%%%%%%%%%%%%%%%%%%%%%%%%%%%%%%%%%%%%%%%%%%%%%%%%%%%
%%%%%%%%%%%%%%%%%%%%%%%%%%%%%%%%%%%%%%%%%%%%%%%%%%%%%%%%%%%%%%%%
\newpage

\bibliography{bib_JMPS1and2}

\begin{thebibliography}{44}
\expandafter\ifx\csname natexlab\endcsname\relax\def\natexlab#1{#1}\fi
\providecommand{\url}[1]{\texttt{#1}}
\providecommand{\href}[2]{#2}
\providecommand{\path}[1]{#1}
\providecommand{\DOIprefix}{doi:}
\providecommand{\ArXivprefix}{arXiv:}
\providecommand{\URLprefix}{URL: }
\providecommand{\Pubmedprefix}{pmid:}
\providecommand{\doi}[1]{\href{http://dx.doi.org/#1}{\path{#1}}}
\providecommand{\Pubmed}[1]{\href{pmid:#1}{\path{#1}}}
\providecommand{\bibinfo}[2]{#2}
\ifx\xfnm\relax \def\xfnm[#1]{\unskip,\space#1}\fi
%Type = Article
\bibitem[{Benoit-Maréchal et~al.(2021)Benoit-Maréchal, Jabbour and
  Triantafyllidis}]{Benoit2020}
\bibinfo{author}{Benoit-Maréchal, L.}, \bibinfo{author}{Jabbour, M.},
  \bibinfo{author}{Triantafyllidis, N.}, \bibinfo{year}{2021}.
\newblock \bibinfo{title}{Revisiting scaling laws for step bunching on vicinal
  surfaces}.
\newblock \bibinfo{journal}{(In preparation)} .
%Type = Article
\bibitem[{Chung and Altman(2002)}]{Chung2002}
\bibinfo{author}{Chung, W.F.}, \bibinfo{author}{Altman, M.S.},
  \bibinfo{year}{2002}.
\newblock \bibinfo{title}{Kinetic length, step permeability, and kinetic
  coefficient asymmetry on the {Si}(111) (7x7) surface}.
\newblock \bibinfo{journal}{Physical Review B} \bibinfo{volume}{66}.
\newblock \DOIprefix\doi{10.1103/physrevb.66.075338}.
%Type = Article
\bibitem[{Dufay et~al.(2007)Dufay, Frisch and Debierre}]{Dufay2007}
\bibinfo{author}{Dufay, M.}, \bibinfo{author}{Frisch, T.},
  \bibinfo{author}{Debierre, J.M.}, \bibinfo{year}{2007}.
\newblock \bibinfo{title}{Role of step-flow advection during
  electromigration-induced step bunching}.
\newblock \bibinfo{journal}{Physical Review B} \bibinfo{volume}{75}.
\newblock \DOIprefix\doi{10.1103/physrevb.75.241304}.
%Type = Article
\bibitem[{Fanion et~al.(2000)Fanion, Fern{\'{a}}ndez and Tallec}]{Fanion2000}
\bibinfo{author}{Fanion, T.}, \bibinfo{author}{Fern{\'{a}}ndez, M.},
  \bibinfo{author}{Tallec, P.L.}, \bibinfo{year}{2000}.
\newblock \bibinfo{title}{Deriving adequate formulations for fluid-structure
  interaction problems: from {ALE} to transpiration}.
\newblock \bibinfo{journal}{Revue Europ{\'{e}}enne des {\'{E}}l{\'{e}}ments
  Finis} \bibinfo{volume}{9}, \bibinfo{pages}{681--708}.
\newblock \DOIprefix\doi{10.1080/12506559.2000.10511481}.
%Type = Article
\bibitem[{Ghez et~al.(1990)Ghez, Cohen and Keller}]{Ghez1990}
\bibinfo{author}{Ghez, R.}, \bibinfo{author}{Cohen, H.G.},
  \bibinfo{author}{Keller, J.B.}, \bibinfo{year}{1990}.
\newblock \bibinfo{title}{Stability of crystals that grow or evaporate by step
  propagation}.
\newblock \bibinfo{journal}{Applied Physics Letters} \bibinfo{volume}{56},
  \bibinfo{pages}{1977--1979}.
\newblock \DOIprefix\doi{10.1063/1.103016}.
%Type = Article
\bibitem[{Ghez et~al.(1993)Ghez, Cohen and Keller}]{Ghez1993}
\bibinfo{author}{Ghez, R.}, \bibinfo{author}{Cohen, H.G.},
  \bibinfo{author}{Keller, J.B.}, \bibinfo{year}{1993}.
\newblock \bibinfo{title}{The stability of growing or evaporating crystals}.
\newblock \bibinfo{journal}{Journal of Applied Physics} \bibinfo{volume}{73},
  \bibinfo{pages}{3685--3693}.
\newblock \DOIprefix\doi{10.1063/1.352928}.
%Type = Phdthesis
\bibitem[{Gillet(2000)}]{Gillet2000}
\bibinfo{author}{Gillet, F.}, \bibinfo{year}{2000}.
\newblock \bibinfo{title}{Dynamique non lin\'{e}aire de surfaces vicinales hors
  de \'{e}quilibre}.
\newblock Ph.D. thesis. Universit\'{e} Joseph Fourier, Grenoble, France.
\newblock \URLprefix \url{https://www.theses.fr/2000GRE10204}.
%Type = Phdthesis
\bibitem[{Guin(2018)}]{Guin2018}
\bibinfo{author}{Guin, L.}, \bibinfo{year}{2018}.
\newblock \bibinfo{title}{Electromechanical couplings and growth instabilities
  in semiconductors}.
\newblock \bibinfo{type}{phdthesis}. Universit\'e Paris-Saclay, \'Ecole
  polytechnique.
\newblock \URLprefix \url{https://www.theses.fr/2018SACLX105}.
%Type = Article
\bibitem[{Guin et~al.(2020)Guin, Jabbour, Shaabani-Ardali,
  Benoit-Mar{\'{e}}chal and Triantafyllidis}]{Guin2020a}
\bibinfo{author}{Guin, L.}, \bibinfo{author}{Jabbour, M.},
  \bibinfo{author}{Shaabani-Ardali, L.},
  \bibinfo{author}{Benoit-Mar{\'{e}}chal, L.},
  \bibinfo{author}{Triantafyllidis, N.}, \bibinfo{year}{2020}.
\newblock \bibinfo{title}{Stability of vicinal surfaces: Beyond the quasistatic
  approximation}.
\newblock \bibinfo{journal}{Physical Review Letters} \bibinfo{volume}{124}.
\newblock \DOIprefix\doi{10.1103/physrevlett.124.036101}.
%Type = Article
\bibitem[{Hata et~al.(1993)Hata, Kawazu, Okano, Ueda and Akiyama}]{Hata1993}
\bibinfo{author}{Hata, K.}, \bibinfo{author}{Kawazu, A.},
  \bibinfo{author}{Okano, T.}, \bibinfo{author}{Ueda, T.},
  \bibinfo{author}{Akiyama, M.}, \bibinfo{year}{1993}.
\newblock \bibinfo{title}{Observation of step bunching on vicinal {GaAs}(100)
  studied by scanning tunneling microscopy}.
\newblock \bibinfo{journal}{Applied Physics Letters} \bibinfo{volume}{63},
  \bibinfo{pages}{1625--1627}.
\newblock \DOIprefix\doi{10.1063/1.110716}.
%Type = Article
\bibitem[{Hibino et~al.(2001)Hibino, Hu, Ogino and Tsong}]{Hibino2001}
\bibinfo{author}{Hibino, H.}, \bibinfo{author}{Hu, C.W.},
  \bibinfo{author}{Ogino, T.}, \bibinfo{author}{Tsong, I.S.T.},
  \bibinfo{year}{2001}.
\newblock \bibinfo{title}{Decay kinetics of two-dimensional islands and holes
  on {Si}(111) studied by low-energy electron microscopy}.
\newblock \bibinfo{journal}{Physical Review B} \bibinfo{volume}{63}.
\newblock \DOIprefix\doi{10.1103/physrevb.63.245402}.
%Type = Article
\bibitem[{Ichimiya et~al.(1996)Ichimiya, Tanaka and Ishiyama}]{Ichimiya1996}
\bibinfo{author}{Ichimiya, A.}, \bibinfo{author}{Tanaka, Y.},
  \bibinfo{author}{Ishiyama, K.}, \bibinfo{year}{1996}.
\newblock \bibinfo{title}{Quantitative measurements of thermal relaxation of
  isolated silicon hillocks and craters on the {Si}(111)-(7x7) surface by
  scanning tunneling microscopy}.
\newblock \bibinfo{journal}{Physical Review Letters} \bibinfo{volume}{76},
  \bibinfo{pages}{4721--4724}.
\newblock \DOIprefix\doi{10.1103/physrevlett.76.4721}.
%Type = Article
\bibitem[{Ishizaki et~al.(1994)Ishizaki, Goto, Kishida, Fukui and
  Hasegawa}]{Ishizaki1994}
\bibinfo{author}{Ishizaki, J.}, \bibinfo{author}{Goto, S.},
  \bibinfo{author}{Kishida, M.}, \bibinfo{author}{Fukui, T.},
  \bibinfo{author}{Hasegawa, H.}, \bibinfo{year}{1994}.
\newblock \bibinfo{title}{Mechanism of multiatomic step formation during
  metalorganic chemical vapor deposition growth of {GaAs} on (001) vicinal
  surface studied by atomic force microscopy}.
\newblock \bibinfo{journal}{Japanese Journal of Applied Physics}
  \bibinfo{volume}{33}, \bibinfo{pages}{721}.
\newblock \DOIprefix\doi{10.1143/JJAP.33.721}.
%Type = Article
\bibitem[{Ishizaki et~al.(1996)Ishizaki, Ohkuri and Fukui}]{Ishizaki1996}
\bibinfo{author}{Ishizaki, J.}, \bibinfo{author}{Ohkuri, K.},
  \bibinfo{author}{Fukui, T.}, \bibinfo{year}{1996}.
\newblock \bibinfo{title}{Simulation and observation of the step bunching
  process grown on {GaAs}(001) vicinal surface by metalorganic vapor phase
  epitaxy}.
\newblock \bibinfo{journal}{Japanese Journal of Applied Physics}
  \bibinfo{volume}{35}, \bibinfo{pages}{1280}.
\newblock \DOIprefix\doi{10.1143/JJAP.35.1280}.
%Type = Article
\bibitem[{Jeong and Williams(1999)}]{Jeong1999}
\bibinfo{author}{Jeong, H.C.}, \bibinfo{author}{Williams, E.D.},
  \bibinfo{year}{1999}.
\newblock \bibinfo{title}{Steps on surfaces: experiment and theory}.
\newblock \bibinfo{journal}{Surface Science Reports} \bibinfo{volume}{34},
  \bibinfo{pages}{171 -- 294}.
\newblock \DOIprefix\doi{10.1016/S0167-5729(98)00010-7}.
%Type = Article
\bibitem[{Johnson et~al.(1997)Johnson, Leung, Birch and Orr}]{Johnson1997}
\bibinfo{author}{Johnson, M.}, \bibinfo{author}{Leung, K.},
  \bibinfo{author}{Birch, A.}, \bibinfo{author}{Orr, B.}, \bibinfo{year}{1997}.
\newblock \bibinfo{title}{Adatom concentration on {GaAs}(001) during
  annealing}.
\newblock \bibinfo{journal}{Journal of Crystal Growth} \bibinfo{volume}{174},
  \bibinfo{pages}{572 -- 578}.
\newblock \DOIprefix\doi{10.1016/S0022-0248(97)00039-0}.
%Type = Article
\bibitem[{Johnson et~al.(1996)Johnson, Leung, Birch, Orr and
  Tersoff}]{Johnson1996}
\bibinfo{author}{Johnson, M.}, \bibinfo{author}{Leung, K.},
  \bibinfo{author}{Birch, A.}, \bibinfo{author}{Orr, B.},
  \bibinfo{author}{Tersoff, J.}, \bibinfo{year}{1996}.
\newblock \bibinfo{title}{Adatom concentration on {GaAs}(001) during mbe
  annealing}.
\newblock \bibinfo{journal}{Surface Science} \bibinfo{volume}{350},
  \bibinfo{pages}{254 -- 258}.
\newblock \DOIprefix\doi{10.1016/0039-6028(95)01110-2}.
%Type = Article
\bibitem[{Kasu and Fukui(1992)}]{Kasu1992}
\bibinfo{author}{Kasu, M.}, \bibinfo{author}{Fukui, T.}, \bibinfo{year}{1992}.
\newblock \bibinfo{title}{Multi-atomic steps on metalorganic chemical vapor
  deposition-grown {GaAs} vicinal surfaces studied by atomic force microscopy}.
\newblock \bibinfo{journal}{Japanese Journal of Applied Physics}
  \bibinfo{volume}{31}, \bibinfo{pages}{L864}.
\newblock \DOIprefix\doi{10.1143/JJAP.31.L864}.
%Type = Article
\bibitem[{Keller et~al.(1993)Keller, Cohen and Merchant}]{Keller1993}
\bibinfo{author}{Keller, J.B.}, \bibinfo{author}{Cohen, H.G.},
  \bibinfo{author}{Merchant, G.J.}, \bibinfo{year}{1993}.
\newblock \bibinfo{title}{The stability of rapidly growing or evaporating
  crystals}.
\newblock \bibinfo{journal}{Journal of Applied Physics} \bibinfo{volume}{73},
  \bibinfo{pages}{3694--3697}.
\newblock \DOIprefix\doi{10.1063/1.352929}.
%Type = Article
\bibitem[{Krug(1997)}]{Krug1997}
\bibinfo{author}{Krug, J.}, \bibinfo{year}{1997}.
\newblock \bibinfo{title}{Origins of scale invariance in growth processes}.
\newblock \bibinfo{journal}{Advances in Physics} \bibinfo{volume}{46},
  \bibinfo{pages}{139--282}.
\newblock \DOIprefix\doi{10.1080/00018739700101498}.
%Type = Inproceedings
\bibitem[{Krug(2005)}]{Krug2005}
\bibinfo{author}{Krug, J.}, \bibinfo{year}{2005}.
\newblock \bibinfo{title}{Introduction to step dynamics and step
  instabilities}, in: \bibinfo{editor}{Voigt, A.} (Ed.),
  \bibinfo{booktitle}{Multiscale Modeling in Epitaxial Growth},
  \bibinfo{publisher}{Birkh{\"a}user Basel}, \bibinfo{address}{Basel}. pp.
  \bibinfo{pages}{69--95}.
\newblock \DOIprefix\doi{10.1007/3-7643-7343-1_6}.
%Type = Article
\bibitem[{Magri et~al.(2014)Magri, Gupta and Rosini}]{Magri2014}
\bibinfo{author}{Magri, R.}, \bibinfo{author}{Gupta, S.K.},
  \bibinfo{author}{Rosini, M.}, \bibinfo{year}{2014}.
\newblock \bibinfo{title}{Step energy and step interactions on the
  reconstructed {GaAs}(001) surface}.
\newblock \bibinfo{journal}{Physical Review B} \bibinfo{volume}{90}.
\newblock \DOIprefix\doi{10.1103/physrevb.90.115314}.
%Type = Article
\bibitem[{Magri et~al.(2016)Magri, Gupta and Rosini}]{Magri2016}
\bibinfo{author}{Magri, R.}, \bibinfo{author}{Gupta, S.K.},
  \bibinfo{author}{Rosini, M.}, \bibinfo{year}{2016}.
\newblock \bibinfo{title}{Erratum: Step energy and step interactions on the
  reconstructed {GaAs}(001) surface [phys. rev. b 90 , 115314 (2014)]}.
\newblock \bibinfo{journal}{Physical Review B} \bibinfo{volume}{94}.
\newblock \DOIprefix\doi{10.1103/physrevb.94.239909}.
%Type = Book
\bibitem[{Michely and Krug(2012)}]{Michely2012}
\bibinfo{author}{Michely, T.}, \bibinfo{author}{Krug, J.},
  \bibinfo{year}{2012}.
\newblock \bibinfo{title}{Islands, mounds and atoms}.
  volume~\bibinfo{volume}{42}.
\newblock \bibinfo{publisher}{Springer Science \& Business Media}.
\newblock \DOIprefix\doi{10.1007/978-3-642-18672-1}.
%Type = Article
\bibitem[{Omi et~al.(2005)Omi, Homma, Tonchev and Pimpinelli}]{Omi2005}
\bibinfo{author}{Omi, H.}, \bibinfo{author}{Homma, Y.},
  \bibinfo{author}{Tonchev, V.}, \bibinfo{author}{Pimpinelli, A.},
  \bibinfo{year}{2005}.
\newblock \bibinfo{title}{New types of unstable step-flow growth on
  {Si}(111)-(7x7) during molecular beam epitaxy: Scaling and universality}.
\newblock \bibinfo{journal}{Physical Review Letters} \bibinfo{volume}{95}.
\newblock \DOIprefix\doi{10.1103/physrevlett.95.216101}.
%Type = Book
\bibitem[{Peyret(2002)}]{peyret2002}
\bibinfo{author}{Peyret, R.}, \bibinfo{year}{2002}.
\newblock \bibinfo{title}{Spectral methods for incompressible viscous flow}.
  volume \bibinfo{volume}{148}.
\newblock \bibinfo{publisher}{Springer-Verlag}.
\newblock \DOIprefix\doi{10.1007/978-1-4757-6557-1}.
%Type = Article
\bibitem[{Pierre-Louis(2003)}]{Pierre-Louis2003}
\bibinfo{author}{Pierre-Louis, O.}, \bibinfo{year}{2003}.
\newblock \bibinfo{title}{Step bunching with general step kinetics: stability
  analysis and macroscopic models}.
\newblock \bibinfo{journal}{Surface Science} \bibinfo{volume}{529},
  \bibinfo{pages}{114 -- 134}.
\newblock \DOIprefix\doi{10.1016/S0039-6028(03)00075-X}.
%Type = Article
\bibitem[{Pimpinelli and Videcoq(2000)}]{Pimpinelli2000}
\bibinfo{author}{Pimpinelli, A.}, \bibinfo{author}{Videcoq, A.},
  \bibinfo{year}{2000}.
\newblock \bibinfo{title}{Novel mechanism for the onset of morphological
  instabilities during chemical vapour epitaxial growth}.
\newblock \bibinfo{journal}{Surface Science} \bibinfo{volume}{445},
  \bibinfo{pages}{L23--L28}.
\newblock \DOIprefix\doi{10.1016/s0039-6028(99)01100-0}.
%Type = Article
\bibitem[{Politi and Krug(2000)}]{Politi2000}
\bibinfo{author}{Politi, P.}, \bibinfo{author}{Krug, J.}, \bibinfo{year}{2000}.
\newblock \bibinfo{title}{Crystal symmetry, step-edge diffusion, and unstable
  growth}.
\newblock \bibinfo{journal}{Surface Science} \bibinfo{volume}{446},
  \bibinfo{pages}{89--97}.
\newblock \DOIprefix\doi{10.1016/s0039-6028(99)01104-8}.
%Type = Article
\bibitem[{Pond(1994)}]{Pond1994}
\bibinfo{author}{Pond, K.}, \bibinfo{year}{1994}.
\newblock \bibinfo{title}{Step bunching and step equalization on vicinal
  {GaAs}(001) surfaces}.
\newblock \bibinfo{journal}{Journal of Vacuum Science {\&} Technology B:
  Microelectronics and Nanometer Structures} \bibinfo{volume}{12},
  \bibinfo{pages}{2689}.
\newblock \DOIprefix\doi{10.1116/1.587232}.
%Type = Article
\bibitem[{Ranguelov et~al.(2017)Ranguelov, Muller, Metois and
  Stoyanov}]{Ranguelov2017}
\bibinfo{author}{Ranguelov, B.}, \bibinfo{author}{Muller, P.},
  \bibinfo{author}{Metois, J.J.}, \bibinfo{author}{Stoyanov, S.},
  \bibinfo{year}{2017}.
\newblock \bibinfo{title}{Step density waves on growing vicinal crystal
  surfaces {\textendash} theory and experiment}.
\newblock \bibinfo{journal}{Journal of Crystal Growth} \bibinfo{volume}{457},
  \bibinfo{pages}{184--187}.
\newblock \DOIprefix\doi{10.1016/j.jcrysgro.2016.06.041}.
%Type = Article
\bibitem[{Ranguelov and Stoyanov(2007)}]{Ranguelov2007}
\bibinfo{author}{Ranguelov, B.}, \bibinfo{author}{Stoyanov, S.},
  \bibinfo{year}{2007}.
\newblock \bibinfo{title}{Evaporation and growth of crystals: Propagation of
  step-density compression waves at vicinal surfaces}.
\newblock \bibinfo{journal}{Physical Review B} \bibinfo{volume}{76}.
\newblock \DOIprefix\doi{10.1103/physrevb.76.035443}.
%Type = Article
\bibitem[{Ranguelov and Stoyanov(2008)}]{Ranguelov2008}
\bibinfo{author}{Ranguelov, B.}, \bibinfo{author}{Stoyanov, S.},
  \bibinfo{year}{2008}.
\newblock \bibinfo{title}{Instabilities at vicinal crystal surfaces:
  Competition between electromigration of adatoms and kinetic memory effect}.
\newblock \bibinfo{journal}{Physical Review B} \bibinfo{volume}{77}.
\newblock \DOIprefix\doi{10.1103/physrevb.77.205406}.
%Type = Article
\bibitem[{Rogilo et~al.(2013)Rogilo, Fedina, Kosolobov, Ranguelov and
  Latyshev}]{Rogilo2013}
\bibinfo{author}{Rogilo, D.I.}, \bibinfo{author}{Fedina, L.I.},
  \bibinfo{author}{Kosolobov, S.S.}, \bibinfo{author}{Ranguelov, B.S.},
  \bibinfo{author}{Latyshev, A.V.}, \bibinfo{year}{2013}.
\newblock \bibinfo{title}{Critical terrace width for two-dimensional nucleation
  during si growth on {Si}(111)-(7x7) surface}.
\newblock \bibinfo{journal}{Physical Review Letters} \bibinfo{volume}{111}.
\newblock \DOIprefix\doi{10.1103/physrevlett.111.036105}.
%Type = Article
\bibitem[{Salmi et~al.(1999)Salmi, Alatalo, Ala-Nissila and
  Nieminen}]{Salmi1999}
\bibinfo{author}{Salmi, M.}, \bibinfo{author}{Alatalo, M.},
  \bibinfo{author}{Ala-Nissila, T.}, \bibinfo{author}{Nieminen, R.},
  \bibinfo{year}{1999}.
\newblock \bibinfo{title}{Energetics and diffusion paths of gallium and arsenic
  adatoms on flat and stepped {GaAs}(001) surfaces}.
\newblock \bibinfo{journal}{Surface Science} \bibinfo{volume}{425},
  \bibinfo{pages}{31--47}.
\newblock \DOIprefix\doi{10.1016/s0039-6028(99)00180-6}.
%Type = Article
\bibitem[{Sekerka(1967)}]{Sekerka1967}
\bibinfo{author}{Sekerka, R.}, \bibinfo{year}{1967}.
\newblock \bibinfo{title}{Application of the time-dependent theory of interface
  stability to an isothermal phase transformation}.
\newblock \bibinfo{journal}{Journal of Physics and Chemistry of Solids}
  \bibinfo{volume}{28}, \bibinfo{pages}{983--994}.
\newblock \DOIprefix\doi{10.1016/0022-3697(67)90215-6}.
%Type = Article
\bibitem[{Shinohara and Inoue(1995)}]{Shinohara1995}
\bibinfo{author}{Shinohara, M.}, \bibinfo{author}{Inoue, N.},
  \bibinfo{year}{1995}.
\newblock \bibinfo{title}{Behavior and mechanism of step bunching during
  metalorganic vapor phase epitaxy of {GaAs}}.
\newblock \bibinfo{journal}{Applied Physics Letters} \bibinfo{volume}{66},
  \bibinfo{pages}{1936--1938}.
\newblock \DOIprefix\doi{10.1063/1.113282}.
%Type = Article
\bibitem[{Slanina et~al.(2005)Slanina, Krug and Kotrla}]{Slanina2005}
\bibinfo{author}{Slanina, F.}, \bibinfo{author}{Krug, J.},
  \bibinfo{author}{Kotrla, M.}, \bibinfo{year}{2005}.
\newblock \bibinfo{title}{Kinetics of step bunching during growth: A minimal
  model}.
\newblock \bibinfo{journal}{Physical Review E} \bibinfo{volume}{71}.
\newblock \DOIprefix\doi{10.1103/physreve.71.041605}.
%Type = Article
\bibitem[{{\v{S}}milauer and Vvedensky(1995)}]{Smilauer1995}
\bibinfo{author}{{\v{S}}milauer, P.}, \bibinfo{author}{Vvedensky, D.D.},
  \bibinfo{year}{1995}.
\newblock \bibinfo{title}{Coarsening and slope evolution during unstable
  spitaxial growth}.
\newblock \bibinfo{journal}{Physical Review B} \bibinfo{volume}{52},
  \bibinfo{pages}{14263--14272}.
\newblock \DOIprefix\doi{10.1103/physrevb.52.14263}.
%Type = Article
\bibitem[{Stewart et~al.(1994)Stewart, Pohland and Gibson}]{Stewart1994}
\bibinfo{author}{Stewart, J.}, \bibinfo{author}{Pohland, O.},
  \bibinfo{author}{Gibson, J.M.}, \bibinfo{year}{1994}.
\newblock \bibinfo{title}{Elastic-displacement field of an isolated surface
  step}.
\newblock \bibinfo{journal}{Phys. Rev. B} \bibinfo{volume}{49},
  \bibinfo{pages}{13848--13858}.
\newblock \DOIprefix\doi{10.1103/PhysRevB.49.13848}.
%Type = Article
\bibitem[{Tejedor et~al.(1998)Tejedor, Allegretti, {\v{S}}milauer and
  Joyce}]{Tejedor1998}
\bibinfo{author}{Tejedor, P.}, \bibinfo{author}{Allegretti, F.},
  \bibinfo{author}{{\v{S}}milauer, P.}, \bibinfo{author}{Joyce, B.},
  \bibinfo{year}{1998}.
\newblock \bibinfo{title}{Temperature-dependent unstable homoepitaxy on vicinal
  {GaAs}(110) surfaces}.
\newblock \bibinfo{journal}{Surface Science} \bibinfo{volume}{407},
  \bibinfo{pages}{82--89}.
\newblock \DOIprefix\doi{10.1016/s0039-6028(98)00149-6}.
%Type = Article
\bibitem[{Tersoff et~al.(1997)Tersoff, Johnson and Orr}]{Tersoff1997}
\bibinfo{author}{Tersoff, J.}, \bibinfo{author}{Johnson, M.D.},
  \bibinfo{author}{Orr, B.G.}, \bibinfo{year}{1997}.
\newblock \bibinfo{title}{Adatom densities on {GaAs}: Evidence for
  near-equilibrium growth}.
\newblock \bibinfo{journal}{Phys. Rev. Lett.} \bibinfo{volume}{78},
  \bibinfo{pages}{282--285}.
\newblock \DOIprefix\doi{10.1103/PhysRevLett.78.282}.
%Type = Article
\bibitem[{Vladimirova et~al.(2001)Vladimirova, Vita and
  Pimpinelli}]{Vladimirova2001}
\bibinfo{author}{Vladimirova, M.}, \bibinfo{author}{Vita, A.D.},
  \bibinfo{author}{Pimpinelli, A.}, \bibinfo{year}{2001}.
\newblock \bibinfo{title}{Dimer diffusion as a driving mechanism of the step
  bunching instability during homoepitaxial growth}.
\newblock \bibinfo{journal}{Physical Review B} \bibinfo{volume}{64}.
\newblock \DOIprefix\doi{10.1103/physrevb.64.245420}.
%Type = Article
\bibitem[{Voigtlander et~al.(1995)Voigtlander, Zinner, Weber and
  Bonzel}]{Voigtlander1995}
\bibinfo{author}{Voigtlander, B.}, \bibinfo{author}{Zinner, A.},
  \bibinfo{author}{Weber, T.}, \bibinfo{author}{Bonzel, H.P.},
  \bibinfo{year}{1995}.
\newblock \bibinfo{title}{Modification of growth kinetics in
  surfactant-mediated epitaxy}.
\newblock \bibinfo{journal}{Physical Review B} \bibinfo{volume}{51},
  \bibinfo{pages}{7583--7591}.
\newblock \DOIprefix\doi{10.1103/physrevb.51.7583}.

\end{thebibliography}
%%%%%%%%%%%%%%%%%%%%%%%%%%%%%%%%%%%%%%%%%%%%%%%%%%%%%%%%%%%%%%%%
%%%%%%%%%%%%%%%%%%%%%%%%%%%%%%%%%%%%%%%%%%%%%%%%%%%%%%%%%%%%%%%%

\end{document}